\documentclass[a4paper,12pt]{article}
\usepackage{epsfig}
\usepackage{amsmath}
\usepackage{amssymb}
\usepackage{amsfonts}
\usepackage{bbm}
\usepackage{cite}
\usepackage[footnotesize]{caption}
\usepackage{graphicx}
\usepackage{mathrsfs}
\usepackage[center,footnotesize,hang]{subfigure}

\def\be{\begin{equation}}
\def\ee{\end{equation}}
\def\beq{\begin{equation}}
\def\eeq{\end{equation}}
\def\bea{\begin{eqnarray}}
\def\eea{\end{eqnarray}}

\def\<{\left\langle}
\def\>{\right\rangle}

\allowdisplaybreaks[2]  \evensidemargin
0cm \oddsidemargin  0cm
\begin{document}
\bibliographystyle{OurBibTeX}
\begin{titlepage}
 \vspace*{-15mm}
\begin{flushright}
\end{flushright}
\vspace*{5mm}
\begin{center}
{ \sffamily \Huge 
Vacuum misalignment corrections to tri-bimaximal mixing and form dominance}
\\[8mm]
Stephen~F.~King\footnote{E-mail:\texttt{king@soton.ac.uk}}
\\[3mm]
{\small\it School of Physics and Astronomy,
University of Southampton,\\
Southampton, SO17 1BJ, U.K.
}\\[1mm]
\end{center}
\vspace*{0.75cm}
\begin{abstract}
\noindent 
Tri-bimaximal neutrino mixing may arise from see-saw models
based on family symmetry which is spontaneously broken by 
flavons with particular vacuum alignments. 
In this paper we derive approximate analytic results which express the deviations from tri-bimaximal neutrino mixing due to vacuum misalignment.
We also relate vacuum misalignment to departures from form dominance,
corresponding to complex deviations from the
real orthogonal $R$ matrix, where such corrections are necessary to allow for successful leptogenesis.
The analytic results show that the 
corrections to tri-bimaximal mixing and form dominance depend
on the pattern of the vacuum misalignment,
with the two effects being uncorrelated.
\end{abstract}
\end{titlepage}
\newpage
\setcounter{footnote}{0}

\section{Introduction}

It is well known that the solar and atmospheric data are
consistent with so-called tri-bimaximal (TB) lepton mixing \cite{HPS}
\footnote{Note that the position of the minus signs in the TB mixing matrix
are phase convention dependent. We have adopted the phase conventions
consistent with the standard PDG parametrisation \cite{Nakamura:2010zzi}. }
,
\begin{eqnarray}
U_{TB} = \left( \begin{array}{rrr}
\frac{2}{\sqrt{6}}   & \frac{1}{\sqrt{3}} & 0 \\
-\frac{1}{\sqrt{6}}  & \frac{1}{\sqrt{3}} & \frac{1}{\sqrt{2}} \\
\frac{1}{\sqrt{6}}  & -\frac{1}{\sqrt{3}} & \frac{1}{\sqrt{2}}
\end{array}
\right). \label{MNS0}
\end{eqnarray}
In the flavour basis (i.e. diagonal charged lepton mass basis), it
has been shown that the TB neutrino mass matrix is invariant under
$S,U$ transformations \cite{Lam:2008rs,Lam:2008sh} 
\be
{M^{\nu}_{TB}}\,= S {M^{\nu}_{TB}} S^T\,=
U {M^{\nu}_{TB}} U^T \ . \label{S} \ee
A very straightforward argument \cite{King:2009ap} shows
that this neutrino flavour symmetry group has only four elements
corresponding to Klein's four-group $Z_2^S \times Z_2^U$
corresponding to the two generators $S,U$.
By contrast the diagonal charged lepton mass
matrix (in this basis) satisfies a diagonal phase
symmetry corresponding to the generator $T$. 
The matrices $S,T,U$ form the three generators of the
group $S_4$ in the triplet representation, while the $A_4$
subgroup is generated by $S,T$. 
This suggests that TB lepton mixing matrix calls for a discrete
non-Abelian family symmetry in nature.
There has been a considerable amount of theoretical work in this
direction \cite{Ma:2007wu,King:2005bj,deMedeirosVarzielas:2005ax,King:2006me,deMedeirosVarzielas:2005qg,deMedeirosVarzielas:2006fc,King:2006np,Altarelli:2006kg,Chen:2009um,Frampton:2004ud,Luhn:2007sy, King:2009mk,Hagedorn:2010th,King:2009qt}.

As discussed in \cite{King:2009ap},
the Klein symmetry of the neutrino mass matrix may originate 
either directly as above or accidentally as an
indirect effect of the family symmetry $G_f$. In such
indirect models the flavons responsible for the neutrino masses break $G_f$
completely so that none of the generators of $G_f$ survive.
The Klein symmetry $Z_2^S \times Z_2^U$ emerges as an accidental symmetry 
due to the appearance of quadratic combinations of flavons in the neutrino sector,
with particular vacuum alignments along the columns of the TB matrix.
This is essentially the approach followed in many existing models in the literature
\cite{King:2005bj,deMedeirosVarzielas:2005ax,King:2006me,deMedeirosVarzielas:2005qg,deMedeirosVarzielas:2006fc,King:2006np}.
In such models there is no compelling reason why the vacuum alignments should 
take this form, and it is quite possible to have alternative vacuum alignments which would
lead to alternative types of mixing which violate the Klein symmetry. 

Recently it has been argued \cite{Abbas:2010jw} that TB mixing may not originate from a family symmetry 
at all, discrete or otherwise, but may be a pure accident. 
The authors of \cite{Abbas:2010jw} explored the experimentally allowed violations of the TB symmetry relations present in the effective neutrino mass matrix $M^{\nu}$ and
found that very strong deviations of the neutrino mass matrix element relations arising from from TB mixing were allowed within current experimental errors on the mixing parameters in $U$. 
We point out that $M^{\nu}$ is comprised of
sums of component matrices $C_i$, weighted by neutrino mass eigenvalues $m_i$,
where the $C_i$ are directly linked to the underlying symmetry.
We shall trace the origin of the observation in \cite{Abbas:2010jw} to leading order zeroes in the
$C_3$ matrix which in indirect or accidental models originates from a flavon aligned along the third
column of the TB matrix which has a zero in the first entry. The observation of \cite{Abbas:2010jw} that
large violations of the TB symmetry relations are allowed then translates into the observation that 
this zero entry will have large (formally infinite) fractional corrections due to any finite correction to the vacuum
alignment. However it has already previously 
been pointed out that these zeroes can be filled in at the leading order without
disturbing the tri-bimaximal predictions for the atmospheric and solar angles \cite{King:2009qt}.
We conclude that these results \cite{King:2009qt,Abbas:2010jw} do not disfavour family symmetry models but do 
show that tri-bimaximal mixing may be insensitive to certain corrections to vacuum alignment. 
This provides a motivation for the present study.

In the remainder of the paper we focus on models based on the type I see-saw
mechanism \cite{Minkowski:1977sc} which explain TB mixing as a consequence of 
spontaneously broken family symmetry. In such models
the Dirac neutrino mass matrix in the diagonal right-handed neutrino mass basis
satisfies the conditions of 
form dominance (FD) \cite{Chen:2009um} at leading order. FD means that, in this basis,
the columns of the Dirac mass matrix are proportional to the columns of the TB matrix.
In practice this is achieved by vacuum alignment of flavon fields. 
In the most natural models
\cite{King:2005bj,deMedeirosVarzielas:2005ax,King:2006me,deMedeirosVarzielas:2005qg,deMedeirosVarzielas:2006fc,King:2006np} there is a separate flavon $\phi_i$ contributing to each of the component
matrices $C_i$, and each of the neutrino masses $m_i$ arises from a separate flavon vacuum expectation
value (VEV), and the TB mixing cannot depend on cancellations involving neutrino masses.
By contrast, in the direct models \cite{Altarelli:2006kg,Chen:2009um,Frampton:2004ud},
the component matrices originate from linear combinations of flavon VEVs \cite{Chen:2009um}.

The way the vacuum alignment is achieved is quite model dependent, but in general the mechanisms
may be classified as being due to D-terms \cite{deMedeirosVarzielas:2005qg,Howl:2009ds}, 
F-terms \cite{Altarelli:2006kg} or
extra dimensional orbifold boundary conditions \cite{Burrows:2010wz}.
Although in principle the desired vacuum alignment of the flavon fields
originates from some high energy family symmetry such as $A_4$, albeit in a model dependent way,
in all cases there will be corrections to the leading order vacuum alignment of flavons.
For example, such corrections can fill in the zeroes,
or violate the equality between different components of a flavon VEV.
In addition the effects of higher order operators can  
allow flavons with a particular alignment to pollute the sector containing at leading order only flavons
of a different alignment. All such effects, which in general lead to a violation of the Klein symmetry, 
will referred to here as ``vacuum misalignment'' since in all cases we are perturbing
away from the forms of vacuum alignment which are known to reproduce TB mixing exactly.
It is worth emphasising that the term ``vacuum misalignment'' as used here
could either refer to a leading order
vacuum alignment (in the case where the original form of vacuum alignment contains a zero) or
a correction to the non-zero components of the 
leading order vacuum alignment, and the results in this paper apply to both situations.

One of the main motivations for this paper is to derive analytic
formulae which relate vacuum misalignments to the deviation from TB mixing.
This paper contains the first analytic results in the literature which relate the effect of general 
vacuum misalignments to the deviations from TB mixing. The value of the
analytic approach is that it enables simple physical insights to be obtained which are not possible with a purely numerical approach, and we illustrate this with some simple examples. 
These examples include (admittedly rather arbitrary) special cases where the vacuum misalignment does not lead to any corrections to TB mixing. However the main value of this paper lies not in the special cases we consider but in the general analytic results which relate vacuum misalignment to deviations from TB mixing, and the only purpose of the examples is to provide simple illustrations of the general results.

It is important to differentiate between two distinct consequences of vacuum misalignment, namely (i) deviations to TB mixing, and (ii) departures from FD, where the two effects are in principle independent of each other.
Apart from presenting analytic formulae describing the first effect (i), we also present a formalism which 
enables the second effect (ii) to be discussed, which is also the first time that such effects have been studied
analytically in the literature. Note that 
nearly all models which give TB mixing using a family symmetry also satisfy the conditions
of FD at the leading order, so the FD approximation for the unperturbed alignment is not restrictive at all
in practice and applies to all of the models in
 \cite{Ma:2007wu,King:2005bj,deMedeirosVarzielas:2005ax,King:2006me,deMedeirosVarzielas:2005qg,deMedeirosVarzielas:2006fc,King:2006np,Altarelli:2006kg,Chen:2009um,Frampton:2004ud,Luhn:2007sy, King:2009mk,Hagedorn:2010th,King:2009qt}, for example, which describe TB mixing.

We stress that both sets of analytic results, i.e. which relate vacuum misalignment to 
both (i) TB mixing corrections 
and (ii) FD corrections, are original results which have not appeared before in the literature. 
Furthermore the analytic results are both useful and physically relevant. 
Firstly the analytic results are useful since in practice some degree of 
vacuum misalignment is always present in realistic models which attempt to describe
TB mixing as the result of a family symmetry.
Secondly such vacuum misalignment will have important physical implications regarding 
neutrino oscillation experiments and leptogenesis.
The physical relevance of the results to precision neutrino oscillation experiments is clear since future
experiments will be sensitive to deviations from TB mixing \cite{Bandyopadhyay:2007kx}, 
and the analytic results
enable such deviations to be related to vacuum misalignment in realistic models, which facilitates
theoretical insights which complement the numerical studies.
The physical relevance of the results to leptogenesis is also clear, 
since the lepton asymmetries vanish exactly in the FD limit where it 
would correspond to a real $R$ matrix \cite{Casas:2001sr} for which leptogenesis vanishes 
\cite{Choubey:2010vs}, as previously observed in particular family symmetry models \cite{Antusch:2006cw}.
The analytical expressions we derive for 
the complex corrections to the real $R$ matrix in terms of the vacuum misalignment
are therefore physically relevant since they allow for non-zero leptogenesis.

In this paper, then, we derive approximate analytic formulae which relate general
vacuum misalignment, as defined above, to the deviations from TB mixing.
We also relate vacuum misalignment to violations of FD via a small complex angle expansion of 
the orthogonal $R$ matrix.
Conventional wisdom says that vacuum misalignment always leads to 
violation of TB mixing and FD, however 
the resulting analytic formulae show that vacuum misalignment may or may not lead to 
deviations from TB mixing and does not necessarily imply violation of FD either,
with the two effects being uncorrelated.
Also, as already discussed above, the recent analyses
hint that TB mixing may be insensitive to vacuum misalignment \cite{King:2009qt,Abbas:2010jw},
and it is interesting to apply our analytic results to  
study this question here, although this is not the main motivation for the paper.
We emphasise that the results here have very general applicability and 
may be applied to all direct or indirect family symmetry models 
based on spontaneously broken family symmetry in order to estimate the deviations from TB neutrino mixing due to vacuum alignment corrections. However, to use the results here, the Dirac mass matrix must be rotated to the
basis in which the charged lepton and right-handed neutrino mass matrices are both diagonal, which is 
automatically the case for the indirect models, at least approximately.
However, in the case of direct models,
the Dirac mass matrix needs to be rotated to the diagonal right-handed neutrino mass basis
before the results can be applied \cite{Chen:2009um}.
Such models should also be formulated in the diagonal charged lepton mass basis,
corresponding to the choice of diagonal $T$ generator basis \cite{Altarelli:2006kg}.
We stress that it is not the goal of this paper
to study the dynamics of vacuum misalignment via some potential or superpotential
of a particular model, as was done for example in \cite{Hagedorn:2010th}.
Instead we are only interested in the effects of vacuum misalignment on TB mixing and FD,
and for this purpose it is sufficient to simply parameterise the misalignment in a particular basis where
the dynamical origin of the misalignment can have a general origin as discussed above.

We remark that TB deviations due to vacuum alignment corrections have previously only
been studied numerically in the framework of the direct $A_4$ models 
\cite{Barry:2010zk}. We also note that the results in this paper are complementary and 
more general than the analytic results in \cite{Antusch:2010tf} which were
confined to sequential dominance (SD) \cite{King:1998jw}, and were derived in a 
completely different way based on a perturbative
diagonalization of the neutrino mass matrix in powers of 
small neutrino mass ratios assuming a hierarchical mass spectrum.
By contrast, our results here are applicable to any pattern of neutrino masses.
Finally we recall that vacuum alignment corrections, though important and in some cases dominant,
are only one of a number of corrections to TB mixing which may arise
in realistic models, the other ones being renormalisation group corrections, canonical normalisation
corrections and charged lepton corrections, but since 
these have all been studied elsewhere \cite{Antusch:2005gp}
they will not be revisited here.

The layout of the remainder of the paper is as follows.
In section 2 we show that the neutrino mass matrix is comprised of a 
sum of component matrices $C_i$ which are closely related to the Klein symmetry.
We show that the observation in \cite{Abbas:2010jw} that large violations in the neutrino mass matrix
may be allowed consistently with current limits on tri-bimaximal deviations is
due to the presence of leading order zeroes in the $C_3$ matrix.
In section 3 we show how TB mixing arises naturally from the type I see-saw mechanism 
if the conditions of FD are satisfied. Working in the diagonal right-handed neutrino mass basis,
perturbations of the Dirac mass matrix (identified with  
vacuum misalignment) are related to deviations from TB mixing via the effective neutrino mass matrix.
We also relate vacuum misalignment to the complex corrections to the real $R$ matrix predicted by FD.
Section 4 summarises and concludes the paper.

\section{The effective neutrino mass matrix}

\subsection{Symmetry and the component matrices}

Let us begin by considering the general case of leptonic mixing.
In the neutrino flavour basis, in which the charged lepton mass
matrix is diagonal and mixing arises from the neutrino
sector, the effective neutrino mass matrix ${M^{\nu}}$, a complex symmetric matrix
containing six phases, may be diagonalised as,
\begin{equation}
U^{\dagger}P_E
{M^{\nu}}P_E U^{\ast}=\mathrm{diag}\,(m_{1}, \; m_{2}, \; m_{3}) \; ,
\label{diag}
\end{equation}
where we find it convenient to work with three complex
neutrino masses $m_i$ and a mixing matrix 
$U$ containing only one Dirac phase,
where $P_E$ is a diagonal phase matrix.
The usual MNS matrix is written in terms of three real and positive neutrino masses $|m_i|$
as $U_{MNS}=U.P_{Maj}$, 
where $P_{Maj}$ contains two Majorana phases, after absorbing the unphysical phases
$P_E$ and an overall phase in the diagonal charged lepton sector. Thus $U$ is the analogue
of the CKM mixing matrix for quarks, involving three mixing angles $\theta_{ij}$ and one phase
$\delta$, in the standard convention.
Given any such mixing matrix $U$, this enables
the neutrino mass matrix
\footnote{${\tilde{M}^{\nu}}$ is loosely referred to as the neutrino mass
matrix in the literature even though the true effective neutrino mass matrix, as determined for example
by the see-saw mechanism, is ${{M}^{\nu}}$.}
 ${\tilde{M}^{\nu}}=P_E{M^{\nu}}P_E$
to be determined in terms of the three complex neutrino masses,
\begin{equation}\label{eq:csd-tbm0}
{\tilde{M}^{\nu}} = U
\mathrm{diag}\,(m_{1}, \; m_{2}, \; m_{3}) \;  U^T
= m_{1} \Phi_{1}{\Phi_{1}}^{T} + m_{2}
\Phi_{2}{\Phi_{2}}^{T} + m_{3} \Phi_{3}{\Phi_{3}}^{T} \; ,
\end{equation}
corresponding to the orthonormal column vectors $\Phi_i$
which are just equal to the columns of $U$, 
\be
U=(\Phi_1, \Phi_2, \Phi_3) \label{columns} 
\ee 
with the
orthonormality relations, \be {\Phi_i}^{\dagger} \Phi_j = \delta_{ij}.
\label{orthonormal} 
\ee 
It is convenient to define the component matrices
\be C_i= \Phi_i {\Phi_i}^T,
\label{component}
 \ee
 in terms of which the neutrino mass
matrix ${\tilde{M}^{\nu}}$ is simply written as a sum, weighted by neutrino masses, 
\be 
{\tilde{M}^{\nu}}= m_1C_1 +
m_2C_2 +  m_3C_3.
\label{sum1}
\ee 

Let us now apply consider the special case of TB mixing, where the columns of $U=U^{TB}$ 
have particularly simple forms which may be written in the standard parametrisation
(the PDG convention with mixing angles given by 
$\sin \theta_{12}=1/\sqrt{3}$, $\sin \theta_{23}=1/\sqrt{2}$) as \cite{Antusch:2007jd},
\begin{equation}
\label{Phi0} {\Phi}_{1}^{TB}=\frac{1}{\sqrt{6}} \left(
\begin{array}{r}
2 \\
-1 \\
1
\end{array}
\right), \ \ {\Phi}_{2}^{TB}=\frac{1}{\sqrt{3}} \left(
\begin{array}{r}
1 \\
1 \\
-1
\end{array}
\right), \ \ {\Phi}_{3}^{TB}=\frac{1}{\sqrt{2}} \left(
\begin{array}{r}
0 \\
1 \\
1
\end{array}
\right),
\end{equation}
where $U_{MNS}^{TB}=U^{TB}.P_{Maj}$.
In the TB example, $S,U$ in Eq.\ref{S} take the particularly simple forms,
\begin{equation}
S = \frac{1}{3} \left(\begin{array}{ccc}
-1& 2  & -2  \\
2  & -1  & -2 \\
-2 & -2 & -1
\end{array}\right), \qquad
U = \left( \begin{array}{rrr}
-1 & 0 & 0 \\
0 & 0 & 1 \\
0 & 1 & 0
\end{array}\right) \ ,\label{SUgens}
\end{equation}
such that the elements $S,U,T$ generate the closed finite group $G_f=S_4$ 
\cite{Lam:2008rs} which may be used as a family
symmetry capable of enforcing TB mixing.

The important point is that the TB symmetry transformations $G$, contained in $G_f$
leave invariant
not only the effective neutrino mass matrix invariant
but also the component matrices $C_i^{TB}$ 
from which the effective
neutrino mass matrix is formed. In natural models it is these component matrices $C_i^{TB}$ which result from the family symmetry $G_f$, with the effective neutrino matrix emerging as a sum of such matrices weighted by neutrino masses $m_i$ which are not predicted by the symmetry. It may happen that a particular element of 
${\tilde{M}^{\nu}}$
is small due to an accidental cancellation in the sum of terms in Eq.\ref{sum1}, but since the masses are not predicted by symmetry the models have nothing to say about this special point. It is precisely such special points that give the largest deviations from the TB mixing relations studied in \cite{Abbas:2010jw}. It is clear that such special accidental points are irrelevant from the perspective of symmetry models. What is relevant for a discussion of the robustness of the symmetry approach is the deviation of the matrix elements of the component matrices
$C_i^{TB}$ from the TB form due to deviations in the mixing parameters from their TB values, as we discuss later.

\subsection{Direct vs Indirect Models}

As discussed in \cite{King:2009ap},
the flavour symmetry of the neutrino mass matrix may originate
from two quite distinct classes of models. The first class of models,
which we call direct models, are based on a family symmetry $G_f=S_4$,
or a closely related family symmetry as discussed below,
some of whose generators are directly preserved in the
lepton sector and are manifested as part of the observed flavour
symmetry. The second class of models, which we call indirect
models, are based on some more general family symmetry $G_f$ which is completely
broken in the neutrino sector, while the observed neutrino flavour
symmetry $Z_2^S \times Z_2^U$ in the neutrino flavour basis emerges as an
accidental symmetry which
is an indirect effect of the family symmetry $G_f$. In such
indirect models the flavons responsible for the neutrino masses break $G_f$
completely so that none of the generators of $G_f$ survive in the
observed flavour symmetry $Z_2^S \times Z_2^U$.

In the direct models, the symmetry of the neutrino mass matrix in
the neutrino flavour basis (which we are calling the neutrino mass matrix for
brevity) is a remnant of the $G_f=S_4$ symmetry of the Lagrangian,
where the generators
$S,U$ are preserved in the neutrino sector, while the diagonal generator
$T$ is preserved in the charged lepton sector.
For direct models, a larger family symmetry $G_f$ which
contains $S_4$ as a subgroup is also possible e.g. $G_f=PSL(2,7)$ \cite{King:2009mk}.
Typically direct models require flavon F-term
vacuum alignment and may include an $SU(5)$ type unification \cite{Altarelli:2006kg}. 
Such minimal $A_4$ models lead to
neutrino mass sum rules between the three masses $m_i$, resulting
in/from a simplified neutrino mass matrix.
$A_4\times SU(5)$ SUSY GUT models are typically constructed in extra dimensions
\cite{Altarelli:2006kg},
where such models in 8D enables vacuum alignment to be elegantly 
achieved by boundary conditions \cite{Burrows:2010wz}.

In the indirect models \cite{King:2009ap} the idea is that the three columns of $U_{TB}$,
$\Phi_i^{TB}$, are promoted to new Higgs fields called
``flavons'', with the
particular vacuum alignments along the directions
$\Phi_i^{TB}$ in Eq.\ref{Phi0} breaking the family symmetry. In the indirect models
the underlying family symmetry of the Lagrangian $G_f$ is
completely broken, and the flavour symmetry of the neutrino mass
matrix $Z_2^S \times Z_2^U$ emerges entirely as an accidental
symmetry, due to the quadratic appearance of such flavons in effective Majorana
Lagrangian which results in a neutrino mass matrix of the desired form in Eq.\ref{eq:csd-tbm0}
\cite{King:2009ap}. 
Such vacuum alignments can be
elegantly achieved using D-term vacuum alignment, which allows
the large classes of discrete family symmetry $G_f$,
namely the $\Delta(3n^2)$ and $\Delta(6n^2)$ groups \cite{King:2009ap}.
We shall discuss an explicit example of an indirect model in subsection~\ref{indirect}.

\subsection{Deviations from tri-bimaximal mixing}
In general, not assuming TB mixing, we can write the neutrino mass matrix as a sum of the component matrices weighted by the neutrino masses:
\begin{equation}
{\tilde{M}^{\nu}}= m_{1} \Phi_{1}{\Phi_{1}}^{T} + m_{2}
\Phi_{2}{\Phi_{2}}^{T} + m_{3} \Phi_{3}{\Phi_{3}}^{T} \; ,
\label{generalMnu}
\end{equation}
where $\Phi_i$ are the orthonormal columns of the mixing matrix 
$U= \left(\Phi_1,\Phi_2,\Phi_3\right)$.
If we are close to the TB case, as current data tells us that we must be, then we can expand the columns of $U$ to lowest order as:
\be
\Phi_i = \Phi_i^{TB}+\Delta \Phi_i.
\label{DeltaPhi}
\ee
Expanding Eqs.\ref{generalMnu} to lowest order in \ref{DeltaPhi},  
\bea
{\tilde{M}^{\nu}} & \approx &  m_{1}[ \Phi_{1}^{TB}{\Phi_{1}^{TB}}^{T} +  \Phi_{1}^{TB}{\Delta \Phi_{1}}^{T} 
+ \Delta \Phi_{1}{\Phi_{1}^{TB}}^{T}] \nonumber \\
& + &  m_{2}[ \Phi_{2}^{TB}{\Phi_{2}^{TB}}^{T} +  \Phi_{2}^{TB}{\Delta \Phi_{2}}^{T} 
+ \Delta \Phi_{2}{\Phi_{2}^{TB}}^{T}] \nonumber \\
& + &  m_{3}[ \Phi_{3}^{TB}{\Phi_{3}^{TB}}^{T} +  \Phi_{3}^{TB}{\Delta \Phi_{3}}^{T} 
+ \Delta \Phi_{3}{\Phi_{3}^{TB}}^{T}].
\label{generalMnuexp}
\eea
 
In the following discussion it is convenient use the expansion about TB mixing
introduced in \cite{King:2007pr},
\be
s_{13} = \frac{r}{\sqrt{2}}, \ \ s_{12} = \frac{1}{\sqrt{3}}(1+s),
\ \ s_{23} = \frac{1}{\sqrt{2}}(1+a),
\label{rsa}
\ee
where the three real parameters
$r,s,a$ describe the deviations of the (r)eactor, (s)olar and
(a)tmospheric angles from their tri-bimaximal values.

The global fits of the conventional mixing angles \cite{GonzalezGarcia:2010er}
can be translated into the
$1\sigma$ ranges:
\begin{eqnarray}
&& 0.07<r<0.21,\ \ -0.05<s<0.003, \ \ 
 -0.09<a<0.04.
\label{rsaexp}
\end{eqnarray}

To first order
in $r,s,a$ the lepton mixing matrix $U$ 
(where as usual $U_{MNS}=U.P_{Maj}$) can be written as \cite{King:2007pr},
\footnote{Other related proposals to parametrize the lepton mixing matrix
have been considered in \cite{Li:2004dn}.}
\begin{eqnarray}
U =
\left( \begin{array}{ccc}
\sqrt{\frac{2}{3}}(1-\frac{1}{2}s)  & \frac{1}{\sqrt{3}}(1+s) & \frac{1}{\sqrt{2}}re^{-i\delta } \\
-\frac{1}{\sqrt{6}}(1+s-a + re^{i\delta })  & \frac{1}{\sqrt{3}}(1-\frac{1}{2}s-a- \frac{1}{2}re^{i\delta })
& \frac{1}{\sqrt{2}}(1+a) \\
\frac{1}{\sqrt{6}}(1+s+a- re^{i\delta })  & -\frac{1}{\sqrt{3}}(1-\frac{1}{2}s+a+ \frac{1}{2}re^{i\delta })
 & \frac{1}{\sqrt{2}}(1-a)
\end{array}
\right),
\label{MNS1}
\end{eqnarray}
from which the deviations of the columns of $U= \left(\Phi_1,\Phi_2,\Phi_3\right)$
in Eq.\ref{DeltaPhi}, namely $\Delta \Phi_i$, may be read off as follows,
\begin{equation}
\label{DeltaPhii} 
\Delta \Phi_1 =
\frac{1}{\sqrt{6}} \left(
\begin{array}{c}
-s \\
-s+a-re^{i\delta } \\
s+a-re^{i\delta } 
\end{array}
\right), 
\Delta \Phi_2 =
\frac{1}{\sqrt{3}} \left(
\begin{array}{c}
s \\
-\frac{1}{2}s-a- \frac{1}{2}re^{i\delta } \\
\frac{1}{2}s-a- \frac{1}{2}re^{i\delta }
\end{array}
\right), 
\Delta \Phi_3 =
\frac{1}{\sqrt{2}} \left(
\begin{array}{c}
re^{-i\delta} \\
a \\
-a
\end{array}
\right).
\end{equation}

It is manifest from Eq.\ref{generalMnuexp}
that the deviations in the component matrices $C_i^{TB}= \Phi_i^{TB} {\Phi_i^{TB}}^T$ from the TB form due to deviations in the mixing parameters from their TB values are proportional to $\Delta \Phi_i$ which are, from 
Eq.\ref{DeltaPhii}, proportional to the mixing deviation parameters $r,s,a$. Thus there is a linear relationship between 
the deviation from the elements of the component matrices and the deviation between the mixing parameters.

\subsection{Example: deviations due to the reactor angle}
Let us now consider as an example the case where TB mixing is only corrected by the presence of a non-zero reactor angle parameterised by the deviation parameter $r$ (with $s=a=0$ in this example) \cite{King:2009qt}. 
This example is interesting since the experimental limit on $r$ is weaker than
on $s,a$ according to Eq.\ref{rsaexp}, and it is also sufficient to understand the 
observations in \cite{Abbas:2010jw}.
In this case, to first order in $r$, from Eqs.\ref{generalMnuexp},\ref{MNS1} we find
that the component matrices $C_i$ which comprise the
neutrino mass matrix can be written as a sum of the TB matrices $C_i^{TB}$ plus a correction
proportional to the reactor parameter $r$,
\bea\label{MTBR2}
C_1=\Phi_{1} \Phi_{1}^{T} & = &  \frac{1}{6}
\left(\begin{array}{rrr}
4 & -2 & 2 \\
-2 & 1 & -1 \\
2 & -1 & 1
\end{array}\right)
- \frac{1}{3}re^{i\delta }
\left(\begin{array}{rrr}
0 & 1 & 1 \\
1 & -1 & 0 \\
1 & 0 & 1
\end{array}\right), \nonumber \\
C_2=\Phi_{2} \Phi_{2}^{T} & = &  \frac{1}{3}
\left(\begin{array}{rrr}
1 & 1 & -1 \\
1 & 1 & -1 \\
-1 & -1 & 1
\end{array}\right)
- \frac{1}{6}re^{i\delta }
\left(\begin{array}{rrr}
0 & 1 & 1 \\
1 & 2 & 0 \\
1 & 0 & -2
\end{array}\right)
, \nonumber \\
C_3=\Phi_{3}\Phi_{3}^{T} & = &  \frac{1}{2}\left(\begin{array}{rrr}
0 & 0 & 0 \\
0 & 1 & 1 \\
0 & 1 & 1
\end{array}\right)
+ \frac{1}{2}re^{-i\delta }
\left(\begin{array}{rrr}
0 & 1 & 1 \\
1 & 0 & 0 \\
1 & 0 & 0
\end{array}\right).
\eea

From Eq.\ref{generalMnu} and the component matrices in Eq.\ref{MTBR2} we may write 
$\tilde{M}^{\nu}$ as the symmetric matrix,
\begin{equation}\label{eq:csd-tbm2}
\tilde{M}^{\nu}=
\left(\begin{array}{ccc}
m_{ee} & m_{e\mu} & m_{e\tau} \\
. & m_{\mu \mu} & m_{\mu \tau} \\
. & . & m_{\tau \tau}
\end{array}\right),
\end{equation}
where,
\begin{eqnarray}\label{abc}
m_{ee} &=& \frac{2}{3}m_1+\frac{1}{3}m_2,\nonumber \\
m_{e\mu} &=& -\frac{1}{3}m_1+\frac{1}{3}m_2
-re^{i\delta }\left(\frac{1}{3}m_1+\frac{1}{6}m_2\right)
+re^{-i\delta }\left(\frac{1}{2}m_3\right)
,\nonumber \\
m_{e\tau} &=& \frac{1}{3}m_1-\frac{1}{3}m_2
-re^{i\delta }\left(\frac{1}{3}m_1+\frac{1}{6}m_2\right)
+re^{-i\delta }\left(\frac{1}{2}m_3\right)
,\nonumber \\
m_{\mu \mu} &=& \frac{1}{6}m_1+\frac{1}{3}m_2 +\frac{1}{2}m_3
+re^{i\delta }\left(\frac{1}{3}m_1-\frac{1}{3}m_2\right)
,\nonumber \\
m_{\tau \tau} &=& \frac{1}{6}m_1+\frac{1}{3}m_2 +\frac{1}{2}m_3
+re^{i\delta }\left(\frac{1}{3}m_1+\frac{1}{3}m_2\right),\nonumber \\
m_{\mu \tau} &=& -\frac{1}{6}m_1-\frac{1}{3}m_2 +\frac{1}{2}m_3.
\end{eqnarray}
In the limit that $r=0$, $\tilde{M}^{\nu}$ reduces to the TB neutrino mass matrix
$\tilde{M}^{\nu}_{TB}$, and the relations $m_{e\mu}=-m_{e\tau}$ and $m_{\mu \mu}=m_{\tau \tau}$ 
and $-m_{\mu \tau}= m_{ee}+m_{e\mu}-m_{\mu \mu}$ emerge
as the characteristic signatures of the TB neutrino mass matrix in the flavour basis,
in the convention for the TB matrix in Eq.\ref{MNS0}. This implies that the origin
of the reactor parameter
$r$ is due to a violation of the family symmetry that would lead to TB mixing.
Following \cite{Abbas:2010jw} we may consider the parameters which signal 
a violation of the TB matrix element relations. For example, in our convention,
we may consider,
\be
\Delta_e = \frac{m_{e\mu} + m_{e\tau}}{m_{e\mu}}. 
\ee
In \cite{Abbas:2010jw} the parameter $\Delta_e$ was shown to suffer very large discrepancies from zero due to a
pole at $m_{e\mu}=0$. The origin of this pole is apparent from the second line of Eq.\ref{abc} where
it is clear that cancellations can occur in the case of a normal hierarchy, for example, where the correction term of order $rm_3$ can compete with the TB term of order $m_2$ for $r$ of order $m_2/m_3$.
From Eq.\ref{MTBR2} it is clear that the component matrix $C_3$ is responsible for this effect since 
$C_3^{TB}$ has zeroes in the first row and column, and thus technically any non-zero value of $r$ will
provide an infinite correction to the symmetry prediction of these elements of the $C_3$ matrix. 
Thus, as emphasised in  \cite{Abbas:2010jw}, this may open the door to alternative approaches to 
neutrino mixing which violate Klein symmetry, especially if $r$ is not much smaller than unity.

We shall see later that the origin of these zeroes in indirect or accidental models is due to 
a flavon aligned along the third column of the TB matrix which has a zero in the first entry. 
If this zero is filled, corresponding to a violation of Klein symmetry,
then this switches on a reactor angle, while preserving the tri-bimaximal
predictions for the solar and atmospheric angles, corresponding to the example discussed in this
subsection. This was first discussed in \cite{King:2009qt} where it was referred to as 
tri-bimaximal reactor mixing.
The symmetry approach should not be abandoned since it 
provides an excellent approximation to and understanding
of the observed near TB mixing which is so far lacking in alternative approaches.
On the contrary, the analyses \cite{Abbas:2010jw,King:2009qt} 
seem to motivate indirect family symmetry models with the
accidental emergence of the Klein symmetry, 
and indicate that tri-bimaximal mixing may be insensitive to relatively
large vacuum misalignment.


\section{The see-saw mechanism}
\subsection{Form Dominance \label{FD}}
We now show how TB mixing can arise at leading order from see-saw models based on
form dominance (FD) \cite{Chen:2009um} which includes most symmetry based models.
To set the notation, recall that,
in the type I see-saw mechanism \cite{Minkowski:1977sc}, the starting point is a heavy right-handed Majorana
neutrino mass matrix
$M_{RR}$ and a Dirac neutrino mass matrix (in the left-right convention) $M_{D}$, with
the light effective left-handed Majorana
neutrino mass matrix $M^{\nu}$ given by the type I see-saw formula
\cite{Minkowski:1977sc},
\begin{equation}\label{eq:meff}
M^{\nu} = M_{D} M_{RR}^{-1} M_{D}^{T}.
\end{equation}
In a basis in which $M_{RR}$ is diagonal with real and positive eigenvalues $M_i$, we may write,
\begin{equation}
M_{RR} = \mbox{diag}(M_1, M_2, M_3)
\end{equation}
and $M_{D}$ may be written in terms of three general column vectors $m_{D1},m_{D2},m_{D3}$,
\begin{equation}
M_{D} = (m_{D1},m_{D2},m_{D3}).
\end{equation}
The see-saw formula then gives,
\begin{equation}
\label{eq:seesawmeff2}
M^{\nu} =
 \frac{m_{D1}m_{D1}^{T}}{M_1} + \frac{m_{D2}m_{D2}^{T}}{M_{2}} +\frac{m_{D3}m_{D3}^{T}}{M_{3}}.
\end{equation}
As first observed in \cite{King:2005bj,Chen:2009um} $M^{\nu}_{TB}$ may be achieved if the columns of the
Dirac mass matrix are aligned along the columns of the TB mixing matrix,
$U^{TB}=(\Phi_1^{TB}, \Phi_2^{TB}, \Phi_3^{TB})$,
\begin{equation}
\label{FD2}
m_{D1}^{TB} = a_1\Phi_1^{TB},\ \
m_{D2}^{TB} = a_2\Phi_2^{TB},\ \
m_{D3}^{TB}  = a_3\Phi_3^{TB},
\end{equation}
where $a_i$ are three complex constants.

Using Eq.\ref{FD2} we see that this
leads to,
\begin{equation}
{M^{\nu}_{TB}}= \frac{a_1^2}{M_1} \Phi_{1}^{TB}{\Phi_{1}^{TB}}^{T} + \frac{a_2^2}{M_2}
\Phi_{2}^{TB}{\Phi_{2}^{TB}}^{T} + \frac{a_3^2}{M_3} \Phi_{3}^{TB}{\Phi_{3}^{TB}}^{T} \; ,
\label{TBMnu}
\end{equation}
diagonalized using Eq.\ref{diag} with 
$U=U^{\mbox{\scriptsize TB}}$ and $P_E= I$  (i.e. zero phases) 
leading to complex neutrino
mass eigenvalues given by $m_1=a_1^2/M_1$, $m_2=a_2^2/M_2$, $m_3=a_3^2/M_3$.
This mechanism
allows a completely general neutrino mass spectrum and, since ${M^{\nu}_{TB}}$
is form diagonalizable (i.e. the mixing angles are independent of the neutrino masses), 
it is referred to as form dominance (FD) \cite{Chen:2009um}.
It is interesting to compare FD to
Constrained Sequential Dominance (CSD) defined in
\cite{King:2005bj}. In CSD a strong hierarchy $|m_1|\ll |m_2| < |m_3|$ is assumed
which enables $m_1$ to be effectively ignored (typically this is achieved by
taking $M_A$ to be very heavy leading to a very light $m_1$) then CSD is defined by
only assuming the second and third conditions in Eq.\ref{FD2} \cite{King:2005bj}.
Thus CSD is seen to be just a special case of FD corresponding to a strong neutrino mass
hierarchy. FD on the other hand is more general
and allows any choice of neutrino masses including
a mild hierarchy, an inverted hierarchy or a quasi-degenerate mass pattern.

In the case of direct symmetry models, for example those in \cite{Altarelli:2006kg},
in the diagonal right-handed neutrino mass basis, each
column vector in Eq.\ref{FD2} corresponds to a linear combination of flavon VEVs,
which requires some mild tuning in order to achieve a mild neutrino mass hierarchy.
To eliminate such tuning one may consider the case that 
each column vector in Eq.\ref{FD2}
arises from a separate flavon VEV, and this possibility,
called natural FD \cite{Chen:2009um}, is realised in
the classes of indirect symmetry models. For example,
if $m_1\ll m_2 < m_3$ then the precise form of $m_{D1}$ becomes
irrelevant, and in this case FD reduces to constrained sequential
dominance (CSD)\cite{King:2005bj}. The CSD mechanism has been
applied in this case to the class of indirect models
with Natural FD based on the family symmetries
$SO(3)$ \cite{King:2005bj,King:2006me} and $SU(3)$
\cite{deMedeirosVarzielas:2005ax}, and their discrete subgroups
\cite{deMedeirosVarzielas:2005qg}. The results here will be most useful for the
indirect models which are naturally expressed in the diagonal right-handed
neutrino mass basis, although the direct models may also be rotated to this basis \cite{Chen:2009um}.

\subsection{Deviations from tri-bimaximal mixing on the see-saw \label{FDdeviations}}

In models based on family symmetry, we have seen that the Dirac mass matrix takes a very special form
in the diagonal right-handed neutrino (and charged lepton) mass basis, namely its columns are proportional
to the columns of the TB mixing matrix $U_{TB}$, as in Eq.\ref{FD2}. This observation is known as FD, since it implies a form diagonalizable neutrino mass matrix. Now we want to consider the effect of deviations $\Delta m_{Di}$, given by,
\be
m_{Di} = m_{Di}^{TB}+\Delta m_{Di},
\label{Delta_mD}
\ee 
and study the resulting deviations from TB mixing corresponding to the mixing matrix being changed to
$U= \left(\Phi_1,\Phi_2,\Phi_3\right)$, where as in Eq.\ref{DeltaPhi},
\be
\Phi_i = \Phi_i^{TB}+\Delta \Phi_i.
\label{DeltaPhi2}
\ee
In this subsection we need determine the linear relation between
$\Delta m_{Di}$ and $\Delta \Phi_i$.  
From the symmetry model building point of view, the $\Delta m_{Di}$ may arise from corrections to vacuum alignment.
From this perspective the results in this subsection provide 
useful relations between TB deviations and vacuum alignment corrections.

Expanding Eqs.\ref{eq:seesawmeff2} to lowest order in the Dirac mass matrix perturbations in Eq.\ref{Delta_mD},  
\be
{M^{\nu}}  \approx   \sum_i \frac{1}{M_i} [ m_{Di}^{TB}{m_{Di}^{TB}}^{T} +  m_{Di}^{TB}{\Delta m_{Di}}^{T} 
+ \Delta m_{Di}{m_{Di}^{TB}}^{T}].
\label{exp_generalMnuexp}
\ee
The first observation is that any deviations $\Delta m_{Di}\propto \Phi_i^{TB}$
will not result in any mixing angle deviations, i.e. $\Delta \Phi_i = 0$ since FD is maintained in this case.
This suggests expanding  $\Delta m_{Di}$ in the TB basis $\Phi_i^{TB}$, 
\be
\Delta m_{Di} = \sum_j \alpha_{ij}\Phi_j^{TB},
\label{DeltamD}
\ee
where $\alpha_{ij}$ are small complex mass parameters, $|\alpha_{ij}|\ll |a_i|$, for all $i,j$,
where $a_i$ are defined by Eq.\ref{FD2}.
Using Eq.\ref{FD2} and Eq.\ref{DeltamD} in Eq.\ref{exp_generalMnuexp},
\be
{M^{\nu}}  \approx   \sum_{i,j} \frac{1}{M_i} [ a_i^2\Phi_{ii}^{TB} + a_i\alpha_{ij}(\Phi_{ij}^{TB} + \Phi_{ji}^{TB}) ].
\label{exp_generalMnuexp2}
\ee
where $\Phi_{ij}^{TB}\equiv \Phi_{i}^{TB}{\Phi_{j}^{TB}}^{T}$.

In order to extract the TB deviation parameters we compare the perturbed neutrino mass matrix in 
Eq.\ref{exp_generalMnuexp2}
to the perturbed neutrino mass matrix in Eq.\ref{generalMnuexp},
repeated below,
\be
{\tilde{M}^{\nu}}  \approx   \sum_i 
m_{i}[ \Phi_{i}^{TB}{\Phi_{i}^{TB}}^{T} +  \Phi_{i}^{TB}{\Delta \Phi_{i}}^{T} 
+ \Delta \Phi_{i}{\Phi_{i}^{TB}}^{T}].
\label{generalMnuexp2}
\ee
The general results for complex leading order neutrino masses $m_i^0=a_i^2/M_i$ are derived in
Appendix~\ref{phases}, with the MNS parameters given in Eq.\ref{solvedrelations} and 
Eq.\ref{mphases} and the neutrino masses in Eq.\ref{neutrinomasssquared}.

In the special case that the leading order neutrino masses are real
(due for example to a real vacuum alignment with $a_i$ real) but allowing 
arbitrary complex vacuum alignment corrections we find from 
Eq.\ref{solvedrelations} of Appendix~\ref{phases} rather compact expressions:
\bea
 s   & \approx &   
 \sqrt{2}\frac{Re\left( \delta m^+_{21}\right)}{m^-_{21}}
    \nonumber \\
     a    & \approx &   
 \sqrt{\frac{2}{3}}\ \frac{Re\left(\delta m^+_{32} \right)}{m^-_{32}}
 -  \frac{1}{\sqrt{3}}\ \frac{Re\left(\delta m^+_{31} \right)}{m^-_{31}}
     \nonumber \\
         re^{-i\delta}    & \approx &   
 \sqrt{\frac{2}{3}}\ \frac{\delta m^+_{32}}{m^-_{32}}
 +  \frac{2}{\sqrt{3}}\  \frac{\delta m^+_{31}}{m^-_{31}}
  \label {solvedrelationsreal}
\eea
where we have written,
\bea
m_i^0&=& \frac{a_i^2}{M_i},  \nonumber \\
m^{\pm}_{ij} &=& m_i^0\pm m_j^0, \nonumber \\
\delta m^+_{ij}&=& \frac{m_i^0\alpha_{ij}}{a_i} +  \frac{m_j^0\alpha_{ji}}{a_j}.
\label{defns}
\eea
From Eq.\ref{neutrinomasssquared} the magnitude of the corrected neutrino masses are:
\be
|m_i|\approx |m_i^0|\left[1+2Re\left(\frac{\alpha_{ii}}{a_i}\right)\right].
\label{neutrinomass}
\ee

\subsection{Vacuum misalignment and deviations from TB mixing \label{indirect}}
In this subsection we shall discuss the application of the results of subsection~\ref{FDdeviations}
to models based on a family symmetry $G_f$.
We shall consider here only an extremely simple example of an indirect model
expressed in the diagonal right-handed neutrino mass basis.
We emphasise that this example is for illustrative purposes only,
and that the results in this paper apply to all models in which TB mixing results from
a family symmetry. For example the results also apply to the  
direct family symmetry models based on $A_4$ when rotated to this basis \cite{Chen:2009um}.

Consider the see-saw Lagrangian in the diagonal charged lepton basis,
\be {\mathcal L}^{Yuk}_N \sim L_i(
y_1\phi_{1}^{i}N^{c}_1 + y_2\phi_{2}^{i}N^{c}_2 + y_3\phi_{3}^{i}N^{c}_3)H
\ , \label{opYuk-6}
\ee
\be {\mathcal L}^{Maj}_{N} \sim M_{1}
N^{c}_1  N^{c}_1 +  M_{2}N^{c}_2  N^{c}_2 + M_{3}N^{c}_3   N^{c}_3
\ , \label{opMaj-6}
\ee
where $y_i$ are Yukawa couplings and these diagonal forms are enforced by additional symmetries.
Since the (CP conjugated)
right-handed neutrinos are family singlets $N^c_i \sim {\bf 1}$, the combination
of family triplet left-handed leptons $L_i \sim {\bf 3}$ and flavons $\phi_i \sim {\bf 3}$ (or $\phi_i \sim
{\bf \overline{3}}$ if the representations are complex) must yield a
singlet of $G_f$. \footnote{Note
that these models are formulated in a basis where the
family indices are trivially summed over. We emphasise again that this particular model with this 
choice of matter and representation content is chosen purely for illustrative purposes.} 
After the see-saw
mechanism takes place, this results in an effective Lagrangian of
the form,
\be {\mathcal L}^{Maj} \sim
L^{}_{} \left( \frac{\phi_{1}\phi_{1}^T}{M_1} +
\frac{\phi_{2}\phi_{2}^T}{M_2} +
\frac{\phi_{3}\phi_{3}^T}{M_3}\right) L^{}_{}HH.
\label{opMaj-7}
\ee
Thus we see the appearance of the quadratic
combinations of flavons which serve to preserve an accidental
neutrino flavour symmetry of the neutrino mass matrix, in the
effective Lagrangian after the see-saw mechanism has taken place.
This is also an example of ``natural FD'' since a separate flavon VEV is responsible
for each physical neutrino mass.

In matrix notation, when the flavons get their VEVs in
the three columns of the Dirac mass matrix
$M_D$ are proportional to the VEVs of the three flavons, 
\be
\label{columns5}
M_D =
(y_1\langle \phi_1\rangle  , y_2\langle \phi_2 \rangle  , y_3\langle \phi_i \rangle )
\equiv (m_{D1},m_{D2},m_{D3}).
\ee 
Thus in the indirect family symmetry models each column of the Dirac mass matrix 
$m_{Di}$ is identified with the VEV of a separate flavon field $\phi_i$, where
\be
\langle \phi_i \rangle \propto m_{Di} = m_{Di}^{TB} + \Delta m_{Di}.
\label{NFD}
\ee

Note that, although we have taken a very specific model here for illustrative purposes, a similar
procedure may be followed for any model in which a general family symmetry $G_f$ leads to TB mixing.
Namely, the general model will have some aligned flavon VEVs which will lead to some Dirac
mass matrix and some heavy Majorana mass matrix in the diagonal charged lepton mass basis.
The Dirac mass matrix of the model in question must then be rotated to the basis in which the 
heavy Majorana mass matrix is diagonal. Then the columns of the Dirac mass matrix in that basis may be
identified with the columns given in Eq.\ref{columns5}. The only difference will be that, in a general model,
the columns of the Dirac mass matrix will not correspond to a unique flavon, but in general will correspond 
to a linear combination of flavons. This makes the analysis of vacuum misalignment more complicated to interpret
than in the simple example considered here, but notwithstanding this complication, the results may be applied to any such model. The main point to note is that all such models satisfy FD
at the leading order \cite{Chen:2009um}, which is the crucial requirement for 
this procedure to be followed.

The leading order vacuum alignment
discussed in \cite{King:2005bj,deMedeirosVarzielas:2005ax,King:2006me,deMedeirosVarzielas:2005qg,deMedeirosVarzielas:2006fc,King:2006np} respects FD with,
\begin{equation}
\label{FD3} 
m_{D1}^{TB}=
\frac{a_1}{\sqrt{6}} \left(
\begin{array}{r}
2 \\
-1 \\
1
\end{array}
\right), 
\ \ m_{D2}^{TB}=
\frac{a_2}{\sqrt{3}} \left(
\begin{array}{r}
1 \\
1 \\
-1
\end{array}
\right), 
\ \ m_{D3}^{TB}=
\frac{a_3}{\sqrt{2}} \left(
\begin{array}{r}
0 \\
1 \\
1
\end{array}
\right).
\end{equation}
The corrections to the leading order vacuum alignment
can be expressed as in Eq.\ref{DeltamD},
\bea
\Delta m_{D1} & \approx & 
\frac{\alpha_{11}}{\sqrt{6}} \left(
\begin{array}{r}
2 \\
-1 \\
1
\end{array}
\right)
+
\frac{\alpha_{12} }{\sqrt{3}} \left(
\begin{array}{r}
1 \\
1 \\
-1
\end{array}
\right)
+
\frac{ \alpha_{13}}{\sqrt{2}} \left(
\begin{array}{r}
0 \\
1 \\
1
\end{array}
\right)
\nonumber \\
\Delta m_{D2} & \approx &
\frac{\alpha_{21}}{\sqrt{6}} \left(
\begin{array}{r}
2 \\
-1 \\
1
\end{array}
\right)
+
\frac{\alpha_{22} }{\sqrt{3}} \left(
\begin{array}{r}
1 \\
1 \\
-1
\end{array}
\right)
+
\frac{ \alpha_{23}}{\sqrt{2}} \left(
\begin{array}{r}
0 \\
1 \\
1
\end{array}
\right)
\nonumber \\
\Delta m_{D3} & \approx &
\frac{\alpha_{31}}{\sqrt{6}} \left(
\begin{array}{r}
2 \\
-1 \\
1
\end{array}
\right)
+
\frac{\alpha_{32} }{\sqrt{3}} \left(
\begin{array}{r}
1 \\
1 \\
-1
\end{array}
\right)
+
\frac{ \alpha_{33}}{\sqrt{2}} \left(
\begin{array}{r}
0 \\
1 \\
1
\end{array}
\right).
\label{DeltaD3}
\eea

The above discussion shows how indirect family symmetry models lead to natural FD at leading order,
since each neutrino mass eigenvalue $m_i$ is associated with 
a particular flavon field $\phi_i$, so no cancellations of flavon VEVs are 
required to generate a particular neutrino mass. In such models the neutrino masses are
free parameters and not predicted by the theory. In the following we consider the case 
of a hierarchical neutrino mass spectrum $|m_1|\ll |m_2| < |m_3|$ where,
since the flavon $\phi_i$ associated with the
neutrino mass $m_i$, it is clear that the flavon $\phi_1$ is irrelevant and may be ignored.
This then reduces to the example of leading order CSD where the dominant flavon $\phi_3$ is responsible
for the atmospheric neutrino mass and mixing angle and the subdominant flavon $\phi_2$ is responsible
for the solar neutrino mass and mixing angle. Including vacuum misalignment, these flavons
have VEVs from Eqs.\ref{NFD},\ref{FD3},\ref{DeltaD3} as follows,
\bea
\langle \phi_2 \rangle &\propto&
\frac{a_2}{\sqrt{3}} \left(
\begin{array}{r}
1 \\
1 \\
-1
\end{array}
\right)
+
\frac{\alpha_{21}}{\sqrt{6}} \left(
\begin{array}{r}
2 \\
-1 \\
1
\end{array}
\right)
+
\frac{\alpha_{22} }{\sqrt{3}} \left(
\begin{array}{r}
1 \\
1 \\
-1
\end{array}
\right)
+
\frac{ \alpha_{23}}{\sqrt{2}} \left(
\begin{array}{r}
0 \\
1 \\
1
\end{array}
\right)
\nonumber \\
\langle \phi_3 \rangle &\propto&
\frac{a_3}{\sqrt{2}} \left(
\begin{array}{r}
0 \\
1 \\
1
\end{array}
\right)
+
\frac{\alpha_{31}}{\sqrt{6}} \left(
\begin{array}{r}
2 \\
-1 \\
1
\end{array}
\right)
+
\frac{\alpha_{32} }{\sqrt{3}} \left(
\begin{array}{r}
1 \\
1 \\
-1
\end{array}
\right)
+
\frac{ \alpha_{33}}{\sqrt{2}} \left(
\begin{array}{r}
0 \\
1 \\
1
\end{array}
\right),
\label{NFD2}
\eea
where $\alpha_{ij}$ and $a_i$ are complex in general with
$|\alpha_{ij}|\ll |a_i|$ so the leading order vacuum alignments are given by the first term
of the right-hand sides, familiar from CSD models \cite{King:2005bj,deMedeirosVarzielas:2005ax,King:2006me,deMedeirosVarzielas:2005qg,deMedeirosVarzielas:2006fc,King:2006np}.
The remaining terms parametrize the vacuum misalignment.

For the case of hierarchical neutrino masses (allowing complex
$\alpha_{ij}$ and $a_i$) the TB mixing deviations parameters are given 
from Eq.\ref{hierarchyrelations2}  of Appendix~\ref{phases}:
\bea
 s   & \approx &  \sqrt{2}Re\left( \frac{\alpha_{21}}{a_2}\right),  
    \nonumber \\
     a    & \approx &   
 \sqrt{\frac{2}{3}}  Re\left( \frac{\alpha_{32}}{a_3}    - \frac{1}{\sqrt{2}} \frac{\alpha_{31}}{a_3}   \right) 
    \nonumber \\
 re^{-i\delta}
    & \approx &   
 \sqrt{\frac{2}{3}}\left( \frac{\alpha_{32}}{a_3} + \sqrt{2}\frac{\alpha_{31}}{a_3}      \right).
  \label {hierarchyrelations3}
\eea

\subsubsection{Preserving the TB solar prediction}
The first observation is that the solar angle deviation parameter $s$ in Eq.\ref{hierarchyrelations3}
is only sensitive to $\phi_2$ vacuum misalignments in Eq.\ref{NFD2}, 
and in particular only those corrections proportional
to $\alpha_{21}$. The solar angle does not care about any $\phi_3$ vacuum misalignments.
Thus the prediction tri-maximal prediction $\sin \theta_{12}=1/\sqrt{3}$ 
corresponding to $s=0$ can be maintained in the presence of any
vacuum alignment corrections such that $\alpha_{21}=0$.
Thus any vacuum alignment correction orthogonal to $\Phi_1^{TB}$
will preserve the TB prediction for the solar angle ($s=0$).
An example of such an alignment is:
\be
\langle \phi_2^{s=0} \rangle \propto
\frac{a_2}{\sqrt{3}} \left(
\begin{array}{r}
1 \\
1 \\
-1
\end{array}
\right)
+
\frac{\alpha_{23}}{\sqrt{2}} \left(
\begin{array}{r}
0 \\
1 \\
1
\end{array}
\right),
\label{solar}
\ee
corresponding to 
$\alpha_{21}=0$ and in addition the optional condition $\alpha_{22}=0$,
chosen to make the misalignment have a simple form.
In these models the tri-bimaximal prediction for the solar angle is therefore relatively robust
in the presence of vacuum alignment corrections, and indeed there is some experimental
support for this observation in Eq.\ref{rsaexp}. If Eq.\ref{solar} is the only vacuum alignment correction then
the atmospheric and reactor angles are
also unchanged and so TB mixing will be preserved with $r=s=a=0$.

\subsubsection{Preserving the TB atmospheric and reactor predictions}
The second observation is that the atmospheric and reactor tri-bimaximal deviation parameters 
in Eq.\ref{hierarchyrelations3} are only sensitive to $\phi_3$ vacuum misalignments in Eq.\ref{NFD2},
and do not care about  $\phi_2$ vacuum misalignments.
Note that the atmospheric and reactor tri-bimaximal deviation parameters 
in Eq.\ref{hierarchyrelations3} do not depend on the parameter $\alpha_{33}$ and hence they are 
insensitive to $\phi_3$ corrections proportional to the leading order 
alignment $\Phi_3^{TB}$, as expected. Since 
the atmospheric and reactor tri-bimaximal deviation parameters 
in Eq.\ref{hierarchyrelations3} depend on different linear combinations of
$\alpha_{32}$ and $\alpha_{31}$ one can envisage corrections for which either
$a=0$ or $r=0$, as we now discuss.

(i)  The case $a=0$ can be achieved for $\alpha_{32}=\alpha_{31}/\sqrt{2}$, 
corresponding to the vacuum misalignment,
\be
\langle \phi_3^{a=0} \rangle \propto
\frac{a_2+\alpha_{33}}{\sqrt{2}} \left(
\begin{array}{c}
re^{-i\delta} \\
1 \\
1
\end{array}
\right),
\label{atm}
\ee
where we have used $re^{-i\delta}=\sqrt{6}\alpha_{32}/a_3$ from Eq.\ref{hierarchyrelations3}.
If in addition $\alpha_{21}=0$, then only the reactor angle and
CP phase $re^{-i\delta}$ are non-zero, and the tri-bimaximal predictions for the
solar and atmospheric angles are both preserved ($s=a=0$).
This was called tri-bimaximal reactor (TBR) mixing in \cite{King:2009qt},
where it was assumed that the only vacuum misalignment was due to Eq.\ref{atm}.
Here we see that additional misalignments such as in Eq.\ref{solar} are also consistent with TBR mixing,
which is a new result.
We emphasise that the vacuum misalignment in Eq.\ref{atm} corresponds to a violation
of Klein symmetry at the leading order, since the first component of $\langle \phi_3^{TB} \rangle$ is zero.
This example corresponds to a leading order vacuum misalignment rather than a correction to a 
vacuum alignment, as discussed in \cite{King:2009qt}. This is related to the observations in
\cite{Abbas:2010jw}, as discussed earlier.

(ii)  The case of zero reactor angle $r=0$ can be achieved for $\alpha_{32}=-\sqrt{2}\alpha_{31}$, 
corresponding to the vacuum misalignment,
\be
\langle \phi_3^{r=0} \rangle \propto
\frac{a_3+\alpha_{33}}{\sqrt{2}} \left(
\begin{array}{r}
0 \\
1 \\
1
\end{array}
\right)
+
\sqrt{\frac{3}{2}}\alpha_{32}
\left(
\begin{array}{c}
0 \\
1 \\
-1
\end{array}
\right).
\label{reactor}
\ee
If in addition $\alpha_{21}=0$, then only the atmospheric angle 
will deviate and the tri-bimaximal predictions for the
solar and reactor angles are both preserved ($s=r=0$).

\subsubsection{Tri-maximal mixing}
One may arrange for the vacuum alignment corrections to lead to the MNS matrix taking the special forms as proposed in the literature (see e.g. \cite{Albright:2010ap} are references therein).
For example, tri-maximal mixing \cite{Grimus:2008tt}, in which the second column of the TB mixing matrix 
is preserved, corresponds to $s=0, a=-\frac{1}{2}r\cos \delta$. From Eq.\ref{hierarchyrelations3}
this be achieved for misalignments with $\alpha_{21}=0$ (giving $s=0$) and $\alpha_{32}=0$ (giving
$a=-\frac{1}{2}r\cos \delta$). An example of a $\phi_3$ misalignment with $\alpha_{32}=0$ is,
\be
\langle \phi_3^{trimax} \rangle \propto
\frac{a_3}{\sqrt{2}} \left(
\begin{array}{c}
0 \\
1 \\
1
\end{array}
\right)
+
\sqrt{\frac{2}{3}}\alpha_{31} \left(
\begin{array}{c}
1 \\
 -1\\
0
\end{array}
\right),
\label{trimax}
\ee
where we have set $\alpha_{33}/\sqrt{2}=-\alpha_{31}/\sqrt{6}$ in order to lead to a simple looking misalignment.

\subsection{Vacuum misalignment and departures from FD}

We have already remarked that violation of FD is welcome since, in the exact FD limit,
corresponding to a real $R$ matrix, leptogenesis asymmetries vanish identically \cite{Choubey:2010vs}.
From this perspective, vacuum misalignment is to be welcomed. However it is not clear
that vacuum misalignment will lead to violation of FD, even though it leads to deviations from TB mixing.
As emphasised in \cite{Chen:2009um}, FD corresponds to the columns of the Dirac mass matrix
$M_{D} = (m_{D1},m_{D2},m_{D3})$
being proportional to the columns of the general MNS matrix
$U=(\Phi_1, \Phi_2, \Phi_3)$. In family symmetry models the MNS matrix is 
identified with the TB mixing matrix and the columns of the Dirac mass matrix then take
simple TB forms which are identified with simple flavon vacuum alignments as discussed in the previous subsection.
Vacuum misalignment will induce departures from the simple Dirac TB forms, resulting in 
$U$ deviating from $U^{TB}$.
However vacuum misalignment will not necessarily induce departures from FD.
The point is that the columns of the corrected Dirac mass matrix may still in principle
be proportional to the columns of the corrected mixing matrix, in which case FD would not be violated
and the leptogenesis asymmetries would remain zero even in the presence of vacuum misalignment.

To investigate this question we recall that 
FD may be expressed in the language of the orthogonal $R$ matrix  \cite{Casas:2001sr} where 
for exact FD the $R$ matrix is a real matrix.
Departures from FD are then signalled by departures of the $R$ matrix from the real matrix. 
In a suitable convention, we can expand the $R$ matrix in a small angle approximation
about the real matrix, and these small angles will be related to vacuum misalignment. 
For many phenomenological applications it is convenient to perform numerical scans over the Dirac mass
matrix parametrized in terms of the orthogonal $R$ matrix, thus it is useful in any case
to be able to have a dictionary between
vacuum misalignment and the $R$ matrix, expanded to leading order in terms of small $R$ matrix angles.

We begin by recalling the derivation of the $R$ matrix in the diagonal
charged lepton and right-handed neutrino mass basis \cite{Casas:2001sr}.
From Eqs.\ref{diag},\ref{eq:meff}, one obtains,
\be
P_{Maj}^{*}U^{\dagger}P_EM_DD_{{M}}^{-1}M_D^TP_EU^*P_{Maj}^{*}=D_k
\,,
\label{start}
\ee
where $D_k$, $D_M$ are diagonal
matrices of positive neutrino mass eigenvalues,
\be
D_k={\rm diag}(|m_1|,|m_2|,|m_3|), \ \ D_M={\rm diag}(M_1,M_2,M_3).
\ee
The $R$ matrix is then defined as,
\be
R=D_{\sqrt{M}}^{-1}M_D^T P_EU^*P_{Maj}^{*}  D_{\sqrt{k}}^{-1}
\, ,
\label{eq:Rmat}
\ee
where from Eq.\ref{start}, 
we see that $R$ is a complex orthogonal matrix $R^TR={I}$.

From Eq.~(\ref{eq:Rmat}) we can write,
\be
P_EM_DD_{\sqrt{M}}^{-1}=UP_{Maj}D_{\sqrt{k}}R^T,
\ee
which shows that the $R$ matrix serves to parametrize $P_EM_D$, 
for fixed values of $U_{MNS}=U.P_{Maj}$, $D_k$ and $D_M$.
It is instructive to expand this equation in terms of the columns of $M_D$ and $U$,
\be
P_E((M_D)_{i1}M_1^{-1/2}, (M_D)_{i2}M_2^{-1/2} , (M_D)_{i3}M_3^{-1/2}) = (U_{i1}m_1^{1/2}, U_{i2}m_2^{1/2} , U_{i3}m_3^{1/2})R^T,
\label{R_explicit}
\ee
reverting again to complex neutrino masses $m_i$.
In the case of FD, where the columns of $M_D$ are proportional to the columns
of $U$, it is apparent that
the orthogonal $R$ matrix is equal to permutations of the unit
matrix with $P_E=I$.
In the convention where the right-handed neutrino of mass $M_i$ is associated with the
physical neutrino of mass $m_i$ in the FD limit we can write $R=I$
\cite{Choubey:2010vs}. In this convention, deviations from FD are then parametrized by a small
$R$ matrix angle expansion. In the standard convention where the right-handed neutrinos
are ordered according to mass $M_1<M_2<M_3$ then this may require a trivial re-ordering of the
columns of the Dirac mass matrix and hence the rows of the $R$ matrix, as is clear from Eq.\ref{R_explicit}
(see also \cite{King:2006hn}).

Taking the transpose of Eq.\ref{R_explicit} we can rewrite this equation in the column vector notation, where
$M_{D} = (m_{D1},m_{D2},m_{D3})$ and
$U=(\Phi_1, \Phi_2, \Phi_3)$,
\be
m_{Di}M_i^{-1/2}=\sum_k R_{ik}m_k^{1/2}\Phi_kP_E^{\dagger}.
\label{Dirac_exp}
\ee
It is clear from Eq.\ref{Dirac_exp} that the $R$ matrix parametrizes an expansion of the columns
of the Dirac mass matrix in the basis of the columns of the {\em perturbed} mixing matrix $U$. 
This is different from our previous approach which was based on an expansion of the 
columns of the 
Dirac mass matrix in the basis of the columns of the {\em unperturbed} TB mixing matrix $U_{TB}$.
Eq.\ref{Dirac_exp} shows that violations of FD are related to
the non-orthogonality of the Dirac columns, since from this equation,
\be
(m_{Dj}^{\dagger}m_{Di})M_j^{-1/2}M_i^{-1/2}=\sum_k R_{ik}|m_k|R_{jk}^*,
\label{Dirac_orthog}
\ee
where the mixing matrices vanish by unitarity.
Eq.\ref{Dirac_orthog} shows that the Dirac columns are orthogonal when the $R$ matrix is diagonal,
and so off-diagonal elements of $R_{ij}$ are associated with
non-orthogonality of the Dirac columns. The since Dirac columns are orthogonal in the FD limit,
we again see that
the violations of FD may thus be parameterised in terms of a small angle expansion of the $R$ matrix,
where Eq.\ref{Dirac_orthog} may be used as the starting point for such an expansion.

The $R$ matrix is a complex orthogonal $3\times 3$ matrix
which can be parameterized in terms of three complex angles
$z_{ij}$ as $R=R_{1}R_{2}R_{3}$ 
where $R_i$ take the form:
\beq
R_{1}=
\left(\begin{array}{ccc}
1 & 0 & 0\\
0 & c_{23} & s_{23}\\
0 & -s_{23} & c_{23}
\end{array}
\right),
R_{2}=
\left(\begin{array}{ccc}
c_{13} & 0 & s_{13}\\
0 & 1 & 0\\
-s_{13} & 0 & c_{13}
\end{array}
\right),
R_{3}=
\left(\begin{array}{ccc}
c_{12} & s_{12} & 0\\
-s_{12} & c_{12} & 0\\
0 & 0 & 1
\end{array}
\right),
\label{RR123}
\eeq
where $s_{ij}=\sin z_{ij}\approx z_{ij}$, $c_{ij}=\cos z_{ij}\approx 1$ in the small
complex angle approximation.

In the small angle approximation, we find the following elements of 
Eq.\ref{Dirac_orthog}:
\bea
(m_{D2}^{\dagger}m_{D1})M_2^{-1/2}M_1^{-1/2}&
\approx &|m_2|z_{12} -|m_1|z_{12}^*+|m_3|z_{23}^*z_{13}\nonumber \\
(m_{D3}^{\dagger}m_{D1})M_3^{-1/2}M_1^{-1/2}&
\approx &|m_3|z_{13} -|m_1|z_{13}^*-|m_2|z_{23}^*z_{12}\nonumber \\
(m_{D3}^{\dagger}m_{D2})M_3^{-1/2}M_2^{-1/2}&
\approx &|m_3|z_{23} -|m_2|z_{23}^*+|m_1|z_{13}^*z_{12}.
\label{Dirac_orthog_elements}
\eea
Expanding the Dirac columns in the TB basis, as in Eqs.\ref{NFD},\ref{FD3},\ref{DeltaD3},
we may evaluate the Dirac matrix elements which appear in Eq.\ref{Dirac_orthog_elements} to first order,
\be
(m_{Dj}^{\dagger}m_{Di})\approx a_j^*a_i\delta_{ji}+\alpha_{ji}^*a_i+a_j^*\alpha_{ij}.
\ee
Eq.\ref{Dirac_orthog_elements} can then be solved to find $R$ matrix complex angles $z_{ij}$.
For example,
in the case of a hierarchical neutrino mass spectrum $|m_1|\ll |m_2| < |m_3|$, Eq.\ref{Dirac_orthog_elements}
may be solved to leading order in the $R$ matrix angles,
\bea
z_{12}&\approx &
(\alpha_{21}^*a_1+a_2^*\alpha_{12})M_2^{-1/2}M_1^{-1/2}|m_2|^{-1}\nonumber \\
z_{13}&\approx &
(\alpha_{31}^*a_1+a_3^*\alpha_{13})M_3^{-1/2}M_1^{-1/2}|m_3|^{-1}\nonumber \\
z_{23}&\approx &
(\alpha_{32}^*a_2+a_3^*\alpha_{23})M_3^{-1/2}M_2^{-1/2}|m_3|^{-1},
\label{smallRangle}
\eea
where the neutrino masses $|m_i|$ are given in Eq.\ref{neutrinomass}, 

Eq.\ref{smallRangle} shows, as expected, that only the off-diagonal vacuum alignment corrections $\alpha_{ij}$ with
$i\neq j$ will lead to non-zero $R$ matrix angles and hence violation of FD.
From Eq.\ref {hierarchyrelations3}, it is seen that, to first order, only $\alpha_{21}$
affects the solar angle deviation from TB mixing, and only $\alpha_{31}$ and $\alpha_{32}$ affect the
atmospheric and reactor deviations from TB mixing. 
Thus it is possible to have vacuum misalignments
which maintain TB mixing,
as discussed in the previous subsection, but which lead to violations of FD
due for example to $\alpha_{12}$, $\alpha_{13}$ and $\alpha_{23}$ being non-zero,
allowing successful leptogenesis.
Alternatively, vacuum misalignment can in principle
lead to deviations from TB mixing with $r,s,a \neq 0$ while maintaining FD with $z_{ij}\approx 0$ due to 
approximate cancellations in Eq.\ref{smallRangle}, $\alpha_{ji}^*a_i+a_j^*\alpha_{ij}\approx 0$.
Clearly having a vacuum misalignment which gives deviations
from TB mixing is not sufficient to guarantee violation of FD and hence successful leptogenesis.

Finally note that, as seen in Eq.\ref{NFD2}, the flavon $\phi_1$,
associated with the right-handed neutrino of mass $M_1$, decouples from the see-saw
mechanism in the limit $m_1\rightarrow 0$, meaning that the TB deviations are 
independent of the alignment of this flavon. However, since $m_1\approx a_1^2/M_1$,
this decoupling may be due to either $a_1\rightarrow 0$ or 
$M_1\rightarrow \infty$. If $M_1\rightarrow \infty$ then Eq.\ref{smallRangle} shows that
$z_{12},z_{13}\rightarrow 0$, which is the two right-handed neutrino limit.
However, if $a_1\rightarrow 0$, with $M_1$ fixed, then Eq.\ref{smallRangle} shows that $z_{12},z_{13}$ 
remain non-zero in addition to $z_{23}$. In this case we can have violations of FD due to 
vacuum misalignment of the flavon $\phi_1$ which is irrelevant for the see-saw mechanism.
This can lead to the $R$ matrix angles $z_{12},z_{13}$ being significantly different from zero,
while maintaining accurate TB mixing, allowing successful leptogenesis
in the framework of CSD models  \cite{Choubey:2010vs}.

\section{Summary and Conclusions}

TB neutrino mixing may arise from see-saw models
based on family symmetry which is spontaneously broken by 
flavons with particular vacuum alignments. 
However in practice some degree of 
vacuum misalignment is always present in realistic models.
In this paper we have derived analytic results which relate such general vacuum misalignment to 
deviations in TB mixing and FD. 
Since the method here only involves inspecting the Dirac mass matrix, 
the results have very general applicability and 
may be applied to all direct or indirect family symmetry models,
including the effects of higher order operators.
However, while the results are readily applicable for indirect models,
the Dirac mass matrix in the direct models
needs to be rotated to the diagonal charged lepton and  
right-handed neutrino mass basis. For example, even if the direct $A_4$ models are formulated
in the diagonal charged lepton basis, the right-handed neutrino mass matrix still needs to be diagonalised
and the Dirac mass matrix correctly identified in this basis before
the results in this paper can be applied. 

The results have important physical implications regarding 
neutrino oscillation experiments and leptogenesis.
Future precision neutrino oscillation experiments will be sensitive to deviations from TB mixing 
and the analytic results presented here
enable such deviations to be related to vacuum misalignment in realistic models.
In simple cases we show that 
certain patterns of vacuum misalignment can preserve TB mixing 
in full or in part with one or more of the TB deviation parameters $r,s,a$ being zero, or can lead 
to tri-maximal mixing where the second columns of the TB matrix is preserved.
The physical relevance of the results to leptogenesis is also clear, 
since the lepton asymmetries vanish exactly in the FD limit where it 
would correspond to a real $R$ matrix.
The analytical expressions in Eq.\ref{smallRangle} for 
the complex corrections to the real $R$ matrix in terms of the vacuum misalignment
are therefore physically relevant since they allow for non-zero leptogenesis.

In conclusion, 
for the classes of family symmetry models studied in the stated approximations, 
the analytic results in this paper provide useful insight into the effects of 
vacuum misalignment on deviations from TB mixing and FD.
The analytic results clearly show how the 
corrections to TB mixing and FD depend
on the pattern of the vacuum misalignment,
with the two effects being uncorrelated.

\section*{Acknowledgements}
We acknowledge partial support from the STFC
Rolling Grant ST/G000557/1.

\section*{Appendix}
\appendix
\section{Effect of vacuum misalignment on TB mixing including phases}\label{phases}
In this appendix we shall give a full derivation of the results relating the deviations
from TB mixing due to vacuum misalignment, including a careful treatment of the phases.
The starting point of the derivation is the 
comparison of the perturbed neutrino mass matrix in Eq.\ref{exp_generalMnuexp2} to the one
in Eq.\ref{generalMnuexp2}.
However, before they can be compared, one must 
take account of the extra phases present in the most general effective neutrino mass matrix 
$M^{\nu}$ (which contains six independent phases) as compared to ${\tilde{M}^{\nu}}$
(which only contains four phases). They are related by ${\tilde{M}^{\nu}}=P_E{M^{\nu}}P_E$.
In the FD limit we saw that $P_E=I$, and close to this limit the phases will be small so that the
diagonal phase matrix is 
approximately equal to the unit matrix. This implies that, to leading order,
Eq.\ref{exp_generalMnuexp2} can be written as
\be
{\tilde{M}^{\nu}} \approx   \sum_{i,j} \frac{1}{M_i} [ a_i^2P_E\Phi_{ii}^{TB}P_E 
+ a_i\alpha_{ij}(\Phi_{ij}^{TB} + \Phi_{ji}^{TB}) ],
\label{exp_generalMnuexp3}
\ee
where the small phases in $P_E$ only modify the leading order terms.
Expanding the phase matrix to first order in the small phases,
\begin{equation}\label{PE}
P_E=
\left(\begin{array}{ccc}
e^{i\delta_1} & 0 & 0 \\
0 & e^{i\delta_2} & 0 \\
0 & 0 & e^{i\delta_3}
\end{array}\right) \approx
I + i\left(\begin{array}{ccc}
\delta_1 & 0 & 0 \\
0 & \delta_2 & 0 \\
0 & 0 & \delta_3
\end{array}\right).
\end{equation}
Eq.\ref{exp_generalMnuexp3} can be written to first order as,
\bea
{\tilde{M}^{\nu}}  
& \approx &
\frac{1}{M_1}\left[\left( a_1^2[1+\frac{i}{3}(4\delta_1+\delta_2+\delta_3)]+2a_1\alpha_{11}     \right) \Phi_{11}^{TB}\right]
  \nonumber \\
&+&
\frac{1}{M_1}\left[\left(a_1\alpha_{12}  + a_1^2\frac{i}{\sqrt{18}}(2\delta_1-\delta_2-\delta_3) \right) 
(\Phi_{12}^{TB} + \Phi_{21}^{TB})\right]
  \nonumber \\
    &+&
\frac{1}{M_1}\left[\left(a_1\alpha_{13}  + a_1^2\frac{i}{\sqrt{12}}(\delta_3-\delta_2) \right) 
(\Phi_{13}^{TB} + \Phi_{31}^{TB})\right]
  \nonumber \\
&+&  
  \frac{1}{M_2}\left[\left( a_2^2[1+\frac{2i}{3}(\delta_1+\delta_2+\delta_3)]+2a_2\alpha_{22}     \right) \Phi_{22}^{TB}\right]
  \nonumber \\
&+&
\frac{1}{M_2}\left[\left(a_2\alpha_{21}  + a_2^2\frac{i}{\sqrt{18}}(2\delta_1-\delta_2-\delta_3) \right) 
(\Phi_{12}^{TB} + \Phi_{21}^{TB})\right]
  \nonumber \\
    &+&
\frac{1}{M_2}\left[\left(a_2\alpha_{23}  + a_2^2\frac{i}{\sqrt{6}}(\delta_2-\delta_3) \right) 
(\Phi_{23}^{TB} + \Phi_{32}^{TB})\right] 
\nonumber \\
&+&  
  \frac{1}{M_3}\left[\left( a_3^2[1+i(\delta_2+\delta_3)]+2a_3\alpha_{33}     \right) \Phi_{33}^{TB}\right]
  \nonumber \\
&+&
\frac{1}{M_3}\left[\left(a_3\alpha_{31}  + a_3^2\frac{i}{\sqrt{12}}(\delta_3-\delta_2) \right) 
(\Phi_{13}^{TB} + \Phi_{31}^{TB})\right]
  \nonumber \\
    &+&
\frac{1}{M_3}\left[\left(a_3\alpha_{32}  + a_3^2\frac{i}{\sqrt{6}}(\delta_2-\delta_3) \right) 
(\Phi_{23}^{TB} + \Phi_{32}^{TB})\right],
  \label {exp_generalMnuexp4}
\eea
where $\Phi_{ij}^{TB}\equiv \Phi_{i}^{TB}{\Phi_{j}^{TB}}^{T}$.

The TB deviation columns $\Delta \Phi_{i}$ in Eq.\ref{DeltaPhii} 
may also be expanded in the TB basis $\Phi_i^{TB}$,
\bea
&& \Delta \Phi_1 = -\frac{s}{\sqrt{2}}\Phi_2^{TB} + \frac{1}{\sqrt{3}}(a-re^{i\delta})\Phi_3^{TB},
 \nonumber \\
&& \Delta \Phi_2 = \frac{s}{\sqrt{2}}\Phi_1^{TB} - \sqrt{\frac{2}{3}}(a+\frac{1}{2}re^{i\delta})\Phi_3^{TB},
 \nonumber \\
&& \Delta \Phi_3 = \frac{2}{\sqrt{6}}(a+\frac{1}{2}re^{-i\delta})\Phi_2^{TB} - \frac{1}{\sqrt{3}}(a-re^{-i\delta})\Phi_1^{TB}. 
\label{DeltaPhiExp}
\eea
Inserting Eq.\ref{DeltaPhiExp} in Eq.\ref{generalMnuexp2},
\bea
{\tilde{M}^{\nu}}  & \approx &
 m_1[ \Phi_{11}^{TB} -  \frac{s}{\sqrt{2}}( \Phi_{12}^{TB}+ \Phi_{21}^{TB})   
+  \frac{1}{\sqrt{3}}(a-re^{i\delta})( \Phi_{13}^{TB}+ \Phi_{31}^{TB})]
  \nonumber \\
&+& m_2[ \Phi_{22}^{TB} +  \frac{s}{\sqrt{2}}( \Phi_{12}^{TB}+ \Phi_{21}^{TB})   
-  \sqrt{\frac{2}{3}}(a+\frac{1}{2}re^{i\delta})( \Phi_{23}^{TB}+ \Phi_{32}^{TB})] \label{generalMnuexp3} \\
&+&  m_3[ \Phi_{33}^{TB} +  
\sqrt{\frac{2}{3}}(a+\frac{1}{2}re^{-i\delta})( \Phi_{23}^{TB}+ \Phi_{32}^{TB})
-  \frac{1}{\sqrt{3}}(a-re^{-i\delta})( \Phi_{13}^{TB}+ \Phi_{31}^{TB})],
  \nonumber 
\eea
where $\Phi_{ij}^{TB}\equiv \Phi_{i}^{TB}{\Phi_{j}^{TB}}^{T}$.

Comparing the coefficients of 
$\Phi_{ij}^{TB}\equiv \Phi_{i}^{TB}{\Phi_{j}^{TB}}^{T} $ in Eq.\ref{exp_generalMnuexp4} to those in 
Eq.\ref{generalMnuexp3} we find the following relations to first order in the small dimensionsless
quantities $r,s,a,\alpha_{ij}/a_i,\delta_i$,
\bea
m_1 & \approx & 
m_1^0\left[ 1+2\frac{\alpha_{11}}{a_1}+ \frac{i}{3}(4\delta_1+\delta_2+\delta_3)\right],
  \nonumber \\
 m_2 & \approx &
m_2^0\left[ 1+2\frac{\alpha_{22}}{a_2}+\frac{2i}{3}(\delta_1+\delta_2+\delta_3)\right], 
   \nonumber \\
 m_3 & \approx &
m_3^0\left[ 1+2\frac{\alpha_{33}}{a_3}+i(\delta_2+\delta_3)\right] , 
   \nonumber \\
  \frac{s}{\sqrt{2}}m^{-}_{21}  & \approx &  
 \frac{im^{+}_{12} }{\sqrt{18}}(2\delta_1-\delta_2-\delta_3)
  +  \delta m^+_{12},  
    \nonumber \\
     \frac{m^0_1}{\sqrt{3}}(a-re^{i\delta}) -   \frac{m^0_3}{\sqrt{3}}(a-re^{-i\delta})
    & \approx &   
 \frac{i m^{+}_{13}}{\sqrt{12}}(\delta_3-\delta_2)
  +   \delta m^+_{13}, 
    \nonumber \\
m^0_3\sqrt{\frac{2}{3}}(a+\frac{1}{2}re^{-i\delta})    -m^0_2\sqrt{\frac{2}{3}}(a+\frac{1}{2}re^{i\delta})   
       & \approx &  
         \frac{i m^{+}_{23}}{\sqrt{6}}(\delta_2-\delta_3)
  +  \delta m^+_{23},
  \label {relations}
\eea
where we have written,
\bea
m_i^0&=& \frac{a_i^2}{M_i},  \nonumber \\
m^{\pm}_{ij} &=& m_i^0\pm m_j^0, \nonumber \\
\delta m^+_{ij}&=& \frac{m_i^0\alpha_{ij}}{a_i} +  \frac{m_j^0\alpha_{ji}}{a_j}.
\eea
Eqs.\ref{relations} may be solved for the three complex neutrino mass eigenvalues 
$m_i=|m_i|e^{i\phi}$, together
with the three real mixing angle deviations $r,s,a$ plus the Dirac oscillation phase $\delta$, 
in terms of the underlying see-saw parameters consisting of the three
real positive heavy right-handed Majorana masses $M_i$, the three complex leading order Dirac 
masses $a_i$, and the 
nine small complex Dirac masses $\alpha_{ij}$. The unphysical phases $\delta_i$
are fixed by the conditions that $r,s,a$ are real. From Eq.\ref{relations} we find the results:

\bea
 s   & \approx &   
 \sqrt{2}Re\left( \frac{\delta m^+_{21}}{m^{-}_{21}} \right)
+ \sqrt{2}  Im\left( \frac{\delta m^+_{21}}{m^-_{21}} \right) \tan \arg \left( \frac{m^{+}_{21}}{m^{-}_{21}} \right),  
    \nonumber \\
     a    & \approx &   
 \sqrt{\frac{2}{3}}\ \frac{Re\left( \frac{\delta m^+_{32}}{m^{+}_{32}} \right)}
 {Re\left( \frac{m^-_{32}}{m^{+}_{32}} \right)}
 -  \sqrt{\frac{1}{3}}\ \frac{Re\left( \frac{\delta m^+_{31}}{m^{+}_{31}} \right)}
 {Re\left( \frac{m^-_{31}}{m^{+}_{31}} \right)}
     \nonumber \\
         r\cos \delta    & \approx &   
 \sqrt{\frac{2}{3}}\ \frac{Re\left( \frac{\delta m^+_{32}}{m^{+}_{32}} \right)}
 {Re\left( \frac{m^-_{32}}{m^{+}_{32}} \right)}
 +  \frac{2}{\sqrt{3}}\ \frac{Re\left( \frac{\delta m^+_{31}}{m^{+}_{31}} \right)}
 {Re\left( \frac{m^-_{31}}{m^{+}_{31}} \right)}
     \nonumber \\
      r\sin \delta    & \approx &   
- \sqrt{\frac{2}{3}}Im\left( \frac{\delta m^+_{32}}{m^{+}_{32}} \right)
+  \sqrt{\frac{2}{3}} Re\left( \frac{\delta m^+_{32}}{m^{+}_{32}} \right) 
\tan \arg \left( \frac{m^{-}_{32}}{m^{+}_{32}} \right)
\nonumber \\
&&  -\frac{2}{\sqrt{3}} Im\left( \frac{\delta m^+_{31}}{m^{+}_{31}} \right)
+ \frac{2}{\sqrt{3}} Re\left( \frac{\delta m^+_{31}}{m^{+}_{31}} \right) 
\tan \arg \left( \frac{m^{-}_{31}}{m^{+}_{31}} \right).
  \label {solvedrelations}
\eea

We write the complex neutrino masses as $m_i=|m_i|e^{i\phi_i}$, and the lowest order
complex masses as $m^0_i=|m^0_i|e^{i\phi^0_i}$.
The magnitude of the neutrino masses are:
\be
|m_i|\approx |m_i^0|\left[1+2Re\left(\frac{\alpha_{ii}}{a_i}\right)\right],
\label{neutrinomasssquared}
\ee
and the phases of the neutrino masses are given by:
\bea
\phi_1 &\approx& \phi^0_1 + 2Im\left(\frac{\alpha_{11}}{a_1}\right)+\frac{1}{3}(4\delta_1+\delta_2+\delta_3)
\nonumber \\
\phi_2 &\approx& \phi^0_2 + 2Im\left(\frac{\alpha_{22}}{a_2}\right)+\frac{2}{3}(\delta_1+\delta_2+\delta_3)
\nonumber \\
\phi_3 &\approx& \phi^0_3 + 2Im\left(\frac{\alpha_{33}}{a_3}\right)+(\delta_2+\delta_3).
\label{mphases}
\eea 
However only the relative neutrino mass phases $\phi_i-\phi_j$ are physical (these are the Majorana phases). 
Only one phase combination appears in the Majorana phases, and this is fixed by the requirement that $s$ is real, which gives,
\be
-\frac{1}{3}(2\delta_1-\delta_2-\delta_3)
\approx 
 \sqrt{2}\ \frac{Im \left( \frac{\delta m^+_{21}}{m^{-}_{21}} \right)}
 {Re\left( \frac{m^+_{21}}{m^{-}_{21}} \right)}.
\label{phasediff}
\ee
For example we find the following Majorana phases,
\bea
\phi_2 - \phi_1
&\approx& \phi^0_2 -  \phi^0_1
 + 2Im\left(\frac{\alpha_{22}}{a_2}\right)
 - 2Im\left(\frac{\alpha_{11}}{a_1}\right)
 +  \sqrt{2}\ \frac{Im \left( \frac{\delta m^+_{21}}{m^{-}_{21}} \right)}
 {Re\left( \frac{m^+_{21}}{m^{-}_{21}} \right)}
\nonumber \\
\phi_3 - \phi_1
&\approx& \phi^0_3 -  \phi^0_1
 + 2Im\left(\frac{\alpha_{33}}{a_3}\right)
 - 2Im\left(\frac{\alpha_{11}}{a_1}\right)
 +  2\sqrt{2}\ \frac{Im \left( \frac{\delta m^+_{21}}{m^{-}_{21}} \right)}
 {Re\left( \frac{m^+_{21}}{m^{-}_{21}} \right)}.
\label{mphases}
\eea 

The results greatly simplify for the case of hierarchical neutrinos,
\be
|m_1|\ll |m_2| <  |m_3|.
\ee
In this case Eq.\ref{relations} simplifies to,
\bea
 s   & \approx &  
 \frac{i }{3}(2\delta_1-\delta_2-\delta_3)
  +  \sqrt{2}\frac{\alpha_{21}}{a_2},  
    \nonumber \\
     a-re^{-i\delta}
    & \approx &   
 \frac{i}{2}(\delta_2-\delta_3) -  \sqrt{3}\frac{\alpha_{31}}{a_3}, 
    \nonumber \\
 a+\frac{r}{2}e^{-i\delta}
    & \approx &   
 \frac{i}{2}(\delta_2-\delta_3) -  \frac{\sqrt{3}}{2}\frac{\alpha_{32}}{a_3}. 
  \label{hierarchyrelations}
\eea
From Eq.\ref{hierarchyrelations} we find,
\bea
 s   & \approx &  \sqrt{2}Re\left( \frac{\alpha_{21}}{a_2}\right),  
    \nonumber \\
     a    & \approx &   
 \sqrt{\frac{2}{3}}  Re\left( \frac{\alpha_{32}}{a_3}    - \frac{1}{\sqrt{2}} \frac{\alpha_{31}}{a_3}   \right) 
    \nonumber \\
 re^{-i\delta}
    & \approx &   
 \sqrt{\frac{2}{3}}\left( \frac{\alpha_{32}}{a_3} + \sqrt{2}\frac{\alpha_{31}}{a_3}      \right), 
  \label{hierarchyrelations2}
\eea
where the phase combination fixed by the requirement of real $s$ is,
\be
-\frac{1}{3}(2\delta_1-\delta_2-\delta_3)
\approx 
 \sqrt{2}\ Im \left( \frac{\alpha_{21}}{a_2} \right).
 \label{hierarchyphasediff}
\ee


\begin{thebibliography}{99}


\bibitem{HPS}
P.~F.~Harrison, D.~H.~Perkins and W.~G.~Scott,
Phys.\ Lett.\ B {\bf 530} (2002) 167 [arXiv:hep-ph/0202074];
P.~F.~Harrison and W.~G.~Scott,
Phys.\ Lett.\ B {\bf 557} (2003) 76 [arXiv:hep-ph/0302025].


\bibitem{Nakamura:2010zzi}
  K.~Nakamura  [Particle Data Group],
  J.\ Phys.\ G {\bf 37} (2010) 075021.
  
\bibitem{Lam:2008rs}
  C.~S.~Lam,
  Phys.\ Rev.\ Lett.\  {\bf 101} (2008) 121602
  [arXiv:0804.2622 [hep-ph]].

\bibitem{Lam:2008sh}
  C.~S.~Lam,
  Phys.\ Rev.\  D {\bf 78} (2008) 073015
  [arXiv:0809.1185 [hep-ph]];
  W.~Grimus, L.~Lavoura and P.~O.~Ludl,
  arXiv:0906.2689 [hep-ph];
  C.~S.~Lam,
  arXiv:0907.2206 [hep-ph].

\bibitem{King:2009ap}
  S.~F.~King and C.~Luhn,
  JHEP {\bf 0910} (2009) 093
  [arXiv:0908.1897 [hep-ph]].



\bibitem{Ma:2007wu}
  E.~Ma and G.~Rajasekaran,
  Phys.\ Rev.\  D {\bf 64} (2001) 113012
  [arXiv:hep-ph/0106291].

\bibitem{King:2005bj}
  S.~F.~King,
  JHEP {\bf 0508}, 105 (2005)
  [arXiv:hep-ph/0506297].

\bibitem{deMedeirosVarzielas:2005ax}
  I.~de Medeiros Varzielas and G.~G.~Ross,
  Nucl.\ Phys.\  B {\bf 733} (2006) 31
  [arXiv:hep-ph/0507176].

\bibitem{King:2006me}
  S.~F.~King and M.~Malinsky,
  JHEP {\bf 0611} (2006) 071
  [arXiv:hep-ph/0608021].

\bibitem{deMedeirosVarzielas:2005qg}
  I.~de Medeiros Varzielas, S.~F.~King and G.~G.~Ross,
  Phys.\ Lett.\  B {\bf 644} (2007) 153
  [arXiv:hep-ph/0512313].

\bibitem{deMedeirosVarzielas:2006fc}
  I.~de Medeiros Varzielas, S.~F.~King and G.~G.~Ross,
  Phys.\ Lett.\  B {\bf 648} (2007) 201
  [arXiv:hep-ph/0607045].

\bibitem{King:2006np}
  S.~F.~King and M.~Malinsky,
  Phys.\ Lett.\  B {\bf 645} (2007) 351
  [arXiv:hep-ph/0610250].

\bibitem{Altarelli:2006kg}
  G.~Altarelli and F.~Feruglio,
  Rev.\ Mod.\ Phys.\  {\bf 82} (2010) 2701
  [arXiv:1002.0211 [hep-ph]];
  G.~Altarelli,
  arXiv:hep-ph/0611117;
G.~Altarelli, F.~Feruglio and Y.~Lin,
  Nucl.\ Phys.\  B {\bf 775} (2007) 31
  [arXiv:hep-ph/0610165];
  T.~J.~Burrows, S.~F.~King,
  Nucl.\ Phys.\  {\bf B835 } (2010)  174-196.
  [arXiv:0909.1433 [hep-ph]];
  G.~Altarelli and F.~Feruglio,
  Nucl.\ Phys.\  B {\bf 741} (2006) 215
  [arXiv:hep-ph/0512103];
  G.~Altarelli and F.~Feruglio,
  Nucl.\ Phys.\  B {\bf 720} (2005) 64
  [arXiv:hep-ph/0504165].


\bibitem{Chen:2009um}
  M.~C.~Chen and S.~F.~King,
  JHEP {\bf 0906} (2009) 072
  [arXiv:0903.0125 [hep-ph]].

\bibitem{Frampton:2004ud}
  P.~H.~Frampton, S.~T.~Petcov and W.~Rodejohann,
  Nucl.\ Phys.\  B {\bf 687} (2004) 31
  [arXiv:hep-ph/0401206];
F.~Plentinger and W.~Rodejohann,
  Phys.\ Lett.\ B {\bf 625} (2005) 264
  [arXiv:hep-ph/0507143];
  T.~Ohlsson and G.~Seidl,
  Nucl.\ Phys.\  B {\bf 643} (2002) 247
  [arXiv:hep-ph/0206087];
  E.~Ma,
  Mod.\ Phys.\ Lett.\  A {\bf 22} (2007) 101
  [arXiv:hep-ph/0610342];
  E.~Ma,
  Mod.\ Phys.\ Lett.\  A {\bf 21} (2006) 1917
  [arXiv:hep-ph/0607056];
  S.~L.~Chen, M.~Frigerio and E.~Ma,
  Nucl.\ Phys.\  B {\bf 724} (2005) 423
  [arXiv:hep-ph/0504181];
  F.~Feruglio, C.~Hagedorn, Y.~Lin and L.~Merlo,
  Nucl.\ Phys.\  B {\bf 775} (2007) 120
  [arXiv:hep-ph/0702194];
  F.~Plentinger and G.~Seidl,
  Phys.\ Rev.\  D {\bf 78} (2008) 045004
  [arXiv:0803.2889 [hep-ph]];
  C.~Csaki, C.~Delaunay, C.~Grojean and Y.~Grossman,
  arXiv:0806.0356 [hep-ph];
  M.-C.~Chen and K.~T.~Mahanthappa,
  Phys.\ Lett.\  B {\bf 652} (2007) 34
  [arXiv:0705.0714 [hep-ph]];
  M.-C.~Chen and K.~T.~Mahanthappa,
  arXiv:0710.2118 [hep-ph];
  R.~N.~Mohapatra and H.~B.~Yu,
  Phys.\ Lett.\  B {\bf 644} (2007) 346
  [arXiv:hep-ph/0610023];
  X.~G.~He,
  Nucl.\ Phys.\ Proc.\ Suppl.\  {\bf 168} (2007) 350
  [arXiv:hep-ph/0612080];
  A.~Aranda,
  arXiv:0707.3661 [hep-ph];
  A.~H.~Chan, H.~Fritzsch and Z.~z.~Xing,
  arXiv:0704.3153 [hep-ph];
  Z.~z.~Xing,
  Phys.\ Lett.\ B {\bf 618} (2005) 141
  [arXiv:hep-ph/0503200];
  Z.~z.~Xing, H.~Zhang and S.~Zhou,
  Phys.\ Lett.\  B {\bf 641} (2006) 189
  [arXiv:hep-ph/0607091];
  S.~K.~Kang, Z.~z.~Xing and S.~Zhou,
  Phys.\ Rev.\  D {\bf 73} (2006) 013001
  [arXiv:hep-ph/0511157];
  S.~Luo and Z.~z.~Xing,
  Phys.\ Lett.\  B {\bf 632} (2006) 341
  [arXiv:hep-ph/0509065];
  M.~Hirsch, E.~Ma, J.~C.~Romao, J.~W.~F.~Valle and A.~Villanova del Moral,
  Phys.\ Rev.\  D {\bf 75} (2007) 053006
  [arXiv:hep-ph/0606082];
  N.~N.~Singh, M.~Rajkhowa and A.~Borah,
  arXiv:hep-ph/0603189;
  X.~G.~He and A.~Zee,
  Phys.\ Lett.\  B {\bf 645} (2007) 427
  [arXiv:hep-ph/0607163];
  N.~Haba, A.~Watanabe and K.~Yoshioka,
  Phys.\ Rev.\ Lett.\  {\bf 97} (2006) 041601
  [arXiv:hep-ph/0603116];
  Z.~z.~Xing,
  Phys.\ Lett.\  B {\bf 533} (2002) 85
  [arXiv:hep-ph/0204049];
  Y.~Lin,
  Nucl.\ Phys.\  B {\bf 813} (2009) 91
  [arXiv:0804.2867 [hep-ph]];
  L.~Yin,
  arXiv:0903.0831 [hep-ph].

\bibitem{Luhn:2007sy}
  C.~Luhn, S.~Nasri and P.~Ramond,
  Phys.\ Lett.\  B {\bf 652} (2007) 27
  [arXiv:0706.2341 [hep-ph]].


\bibitem{King:2009mk}
  S.~F.~King and C.~Luhn,
  arXiv:0905.1686 [hep-ph].




\bibitem{Hagedorn:2010th}
  C.~Hagedorn, S.~F.~King, C.~Luhn,
  JHEP {\bf 1006 } (2010)  048.
  [arXiv:1003.4249 [hep-ph]].

\bibitem{King:2009qt}
  S.~F.~King,
  Phys.\ Lett.\ B {\bf 675} (2009) 347
  [arXiv:0903.3199 [hep-ph]].

\bibitem{Abbas:2010jw}
  M.~Abbas and A.~Y.~Smirnov,
  Phys.\ Rev.\  D {\bf 82} (2010) 013008
  [arXiv:1004.0099 [hep-ph]].

\bibitem{Minkowski:1977sc}
  P.~Minkowski,
  Phys.\ Lett.\  B {\bf 67} (1977) 421.

\bibitem{Howl:2009ds}
  R.~Howl and S.~F.~King,
  Phys.\ Lett.\  B {\bf 687} (2010) 355
  [arXiv:0908.2067 [hep-ph]].

\bibitem{Burrows:2010wz}
  T.~Kobayashi, Y.~Omura and K.~Yoshioka,
  Phys.\ Rev.\  D {\bf 78} (2008) 115006
  [arXiv:0809.3064 [hep-ph]];
  T.~J.~Burrows, S.~F.~King,
  Nucl.\ Phys.\  {\bf B842 } (2011)  107-121.
  [arXiv:1007.2310 [hep-ph]].

\bibitem{Bandyopadhyay:2007kx}
  A.~Bandyopadhyay {\it et al.}  [ISS Physics Working Group],
  Rept.\ Prog.\ Phys.\  {\bf 72} (2009) 106201
  [arXiv:0710.4947 [hep-ph]].

\bibitem{Casas:2001sr}
  J.~A.~Casas and A.~Ibarra,
  Nucl.\ Phys.\ B {\bf 618} (2001) 171
  [arXiv:hep-ph/0103065].

\bibitem{Choubey:2010vs}
  S.~Choubey, S.~F.~King and M.~Mitra,
  Phys.\ Rev.\  D {\bf 82} (2010) 033002
  [arXiv:1004.3756 [hep-ph]].
  
\bibitem{Antusch:2006cw}
  S.~Antusch, S.~F.~King and A.~Riotto,
  JCAP {\bf 0611} (2006) 011
  [arXiv:hep-ph/0609038];
  S.~F.~King,
  Nucl.\ Phys.\  B {\bf 786}, 52 (2007)
  [arXiv:hep-ph/0610239];
  E.~E.~Jenkins and A.~V.~Manohar,
  Phys.\ Lett.\  B {\bf 668}, 210 (2008)
  [arXiv:0807.4176 [hep-ph]];
  E.~Bertuzzo, P.~Di Bari, F.~Feruglio and E.~Nardi,
  JHEP {\bf 0911}, 036 (2009)
  [arXiv:0908.0161 [hep-ph]];
  D.~Aristizabal Sierra, F.~Bazzocchi, I.~de Medeiros Varzielas, L.~Merlo and S.~Morisi,
  Nucl.\ Phys.\  B {\bf 827}, 34 (2010)
  [arXiv:0908.0907 [hep-ph]];
  R.~G.~Felipe and H.~Serodio,
  arXiv:0908.2947 [hep-ph];
  D.~Aristizabal Sierra, F.~Bazzocchi, I.~de Medeiros Varzielas, L.~Merlo and S.~Morisi,
  Nucl.\ Phys.\  B {\bf 827}, 34 (2010)
  [arXiv:0908.0907 [hep-ph]];
  E.~Bertuzzo, P.~Di Bari, F.~Feruglio and E.~Nardi,
  JHEP {\bf 0911}, 036 (2009)
  [arXiv:0908.0161 [hep-ph]].
  
 
  

\bibitem{Barry:2010zk}
  M.~Honda, M.~Tanimoto,
  Prog.\ Theor.\ Phys.\  {\bf 119 } (2008)  583-598.
  [arXiv:0801.0181 [hep-ph]];
    J.~Barry and W.~Rodejohann,
  Phys.\ Rev.\  D {\bf 81} (2010) 093002
  [Erratum-ibid.\  D {\bf 81} (2010) 119901]
  [arXiv:1003.2385 [hep-ph]].
  
 
  
\bibitem{Antusch:2010tf}
  S.~Antusch, S.~Boudjemaa and S.~F.~King,
  JHEP {\bf 1009} (2010) 096
  [arXiv:1003.5498 [hep-ph]].
  
\bibitem{King:1998jw}
S.~F.~King,
Phys.\ Lett.\ B {\bf 439} (1998) 350 [arXiv:hep-ph/9806440];
S.~F.~King,
Nucl.\ Phys.\ B {\bf 562} (1999) 57 [arXiv:hep-ph/9904210];
S.~F.~King,
Nucl.\ Phys.\ B {\bf 576} (2000) 85 [arXiv:hep-ph/9912492];
S.~F.~King,
JHEP {\bf 0209} (2002) 011 [arXiv:hep-ph/0204360];
  S.~Antusch and S.~F.~King,
  New J.\ Phys.\  {\bf 6} (2004) 110
  [arXiv:hep-ph/0405272].
  
  
\bibitem{Antusch:2005gp}
  S.~Antusch, J.~Kersten, M.~Lindner, M.~Ratz and M.~A.~Schmidt,
  JHEP {\bf 0503} (2005) 024
  [arXiv:hep-ph/0501272];
  P.~H.~Frampton, S.~T.~Petcov, W.~Rodejohann,
  Nucl.\ Phys.\  {\bf B687 } (2004)  31-54.
  [hep-ph/0401206];
  Z.~-z.~Xing,
  Phys.\ Lett.\  {\bf B533 } (2002)  85-93.
  [hep-ph/0204049];
  S.~Antusch, S.~F.~King and M.~Malinsky,
  Nucl.\ Phys.\  B {\bf 820} (2009) 32
  [arXiv:0810.3863 [hep-ph]];
  S.~Boudjemaa and S.~F.~King,
  Phys.\ Rev.\  D {\bf 79} (2009) 033001
  [arXiv:0808.2782 [hep-ph]];
  S.~Antusch, S.~F.~King and M.~Malinsky,
  JHEP {\bf 0805} (2008) 066
  [arXiv:0712.3759 [hep-ph]];
  S.~Antusch, S.~F.~King and M.~Malinsky,
  Phys.\ Lett.\  B {\bf 671} (2009) 263
  [arXiv:0711.4727 [hep-ph]];
  S.~F.~King,
  Phys.\ Lett.\  B {\bf 659} (2008) 244
  [arXiv:0710.0530 [hep-ph]];
  S.~Antusch, P.~Huber, S.~F.~King and T.~Schwetz,
  JHEP {\bf 0704} (2007) 060
  [arXiv:hep-ph/0702286];
  S.~Antusch and S.~F.~King,
  Phys.\ Lett.\  B {\bf 631} (2005) 42
  [arXiv:hep-ph/0508044].
  
  
 
  

\bibitem{Antusch:2007jd}
  S.~Antusch, L.~E.~Ibanez and T.~Macri,
  JHEP {\bf 0709} (2007) 087
  [arXiv:0706.2132 [hep-ph]].

\bibitem{GonzalezGarcia:2010er}
  M.~C.~Gonzalez-Garcia, M.~Maltoni and J.~Salvado,
  JHEP {\bf 1004} (2010) 056
  [arXiv:1001.4524 [Unknown]].

\bibitem{King:2007pr}
  S.~F.~King,
  Phys.\ Lett.\  B {\bf 659} (2008) 244
  [arXiv:0710.0530 [hep-ph]].
  
\bibitem{Li:2004dn}
  Z.~z.~Xing,
  Phys.\ Lett.\  B {\bf 533} (2002) 85
  [arXiv:hep-ph/0204049];
  N.~Li and B.~Q.~Ma,
  Phys.\ Rev.\  D {\bf 71} (2005) 017302
  [arXiv:hep-ph/0412126];
Talk by S.~Parke at WIN'05,
20th International Workshop on Weak Interactions and Neutrinos,
European Cultural Center, Delphi, Greece, June 6-11, 2005,
http://conferences.phys.uoa.gr/win05/ ;
Talk by P.~F.~Harrison, ``Deviations from Tri-bimaximal Mixing'',
Rutherford Appleton Laboratory, U.K., April 24 - 28, 2006,
http://www.hep.ph.ic.ac.uk/uknfic/iss0406/physics.html ;
  S.~Pakvasa, W.~Rodejohann and T.~J.~Weiler,
  Phys.\ Rev.\ Lett.\  {\bf 100} (2008) 111801
  [arXiv:0711.0052 [hep-ph]].

 
\bibitem{Albright:2010ap}
  C.~H.~Albright, A.~Dueck and W.~Rodejohann,
  arXiv:1004.2798 [hep-ph].

\bibitem{Grimus:2008tt}
  W.~Grimus and L.~Lavoura,
  JHEP {\bf 0809} (2008) 106
  [arXiv:0809.0226 [hep-ph]].
  
\bibitem{King:2006hn}
  S.~F.~King,
  Nucl.\ Phys.\  B {\bf 786} (2007) 52
  [arXiv:hep-ph/0610239].
  
  










\end{thebibliography}
\end{document}